\documentclass[10pt
,aps 
,prx
,amsmath
,amssymb
,amsfonts
,superscriptaddress
,twocolumn
,notitlepage
,longbibliography
,nofootinbib]{revtex4-2}
\usepackage{comment}
\usepackage{lipsum} 
\usepackage{mathtools}
\usepackage{amsmath,amssymb}
\usepackage{physics}
\usepackage[caption=false]{subfig}
\usepackage{graphicx}
\usepackage{dcolumn}
\usepackage{xcolor}
\usepackage[normalem]{ulem} 
\usepackage[colorlinks=true,linkcolor=blue,urlcolor=black,citecolor=blue,bookmarksopen=true]{hyperref}
\usepackage{mathrsfs}

\usepackage{dsfont}
\usepackage{mathtools}
\usepackage{bbm,bm}
\usepackage{mathdots}
\usepackage{amssymb}
\usepackage{amsfonts}
\newcommand{\cmmnt}[1]{}

\newcommand{\colvec}[2][.8]{%
  \scalebox{#1}{%
    \renewcommand{\arraystretch}{.8}%
    $\begin{pmatrix}#2\end{pmatrix}$%
  }
}
\begin{document}
\title{Time-crystalline behavior in central-spin models with Heisenberg interactions} 

\author{Rafail Frantzeskakis}
\affiliation{Department of Physics, University of Crete, Heraklion, 71003, Greece}

\author{John Van Dyke}
\affiliation{Department of Physics, Virginia Tech, Blacksburg, VA 24061, USA}
\affiliation{Virginia Tech Center for Quantum Information Science and Engineering, Blacksburg, VA 24061, USA}

\author{Leon Zaporski}
\affiliation{Cavendish Laboratory, University of Cambridge,
JJ Thomson Avenue, Cambridge, CB3 0HE, United Kingdom}

\author{Dorian A. Gangloff}
\affiliation{Department of Engineering Science, University of Oxford, Oxford, OX1 3PJ, United Kingdom}

\author{Claire Le Gall}
\affiliation{Microsoft Research, Cambridge, CB1 2FB, United Kingdom}

\author{Mete Atat\"ure}
\affiliation{Cavendish Laboratory, University of Cambridge,
JJ Thomson Avenue, Cambridge, CB3 0HE, United Kingdom}

\author{Sophia E. Economou}
\affiliation{Department of Physics, Virginia Tech, Blacksburg, VA 24061, USA}
\affiliation{Virginia Tech Center for Quantum Information Science and Engineering, Blacksburg, VA 24061, USA}

\author{Edwin Barnes}
\affiliation{Department of Physics, Virginia Tech, Blacksburg, VA 24061, USA}
\affiliation{Virginia Tech Center for Quantum Information Science and Engineering, Blacksburg, VA 24061, USA}
\begin{abstract}
Time-crystalline behavior has been predicted and observed in quantum central-spin systems with periodic driving and Ising interactions. Here, we theoretically show that it can also arise in central-spin systems with Heisenberg interactions. We present two methods to achieve this: application of a sufficiently large Zeeman splitting on the central spin compared to the satellite spins, or else by applying additional pulses to the central spin every Floquet period. In both cases, we show that the system exhibits a subharmonic response in spin magnetizations in the presence of disorder for both pure Heisenberg and XXZ interactions. Our results pertain to any XXZ central-spin system, including hyperfine-coupled electron-nuclear systems in quantum dots or color centers.
\end{abstract}
\maketitle

\section{Introduction}
Spontaneous symmetry breaking has a long history in condensed matter and high-energy physics ~\cite{van2007spontaneous,beekman2019introduction}. In the past decade, time crystals have attracted particular attention, both theoretically and in experiments. On the theoretical side, time crystals enrich the class of nonequilibrium phases of matter, and have close ties to questions about many-body localization and thermalization in quantum systems~\cite{Nandkishore2015,ponte2015many,abanin2019colloquium,khemani2019briefhistory,morningstar2022avalanches}. In a discrete time crystal, the time translation symmetry of a periodic Hamiltonian $H(t+T) = H(t)$ is broken, and expectation values of certain observables exhibit a subharmonic response~\cite{sachamodelling2015,Khemani2016,else2016floquet}. Several different routes leading to discrete time translation symmetry breaking have been extensively studied, including many-body localization in the presence of strong disorder~\cite{else2016floquet,sacha2017time,yao2017discrete,ippoliti2021many,liu2023discrete} and prethermalization, which does not rely on disorder~\cite{russomanno2017floquet,2019angelakiswithoutdis,zhu2019dicke,fan2020discrete,collado2021emergent,collado2023dynamical}. The initial theoretical investigations led to many experimental realizations in different physical platforms such as trapped ions~\cite{zhang2017observation,kyprianidis2021observation}, solid-state spin ensembles~\cite{choi2017observation,rovny2018observation,o2020signatures,randall2021many,beatrez2023critical}, ultracold atoms~\cite{smits2018observation,hannaford2022condensed}, superconducting qubits~\cite{zhang2021observation,xu2021realizing, Mi2022,frey2022realization}, and magnons~\cite{autti2018observation,magnon2021,autti2022nonlinear}.  Apart from closed  systems, there also exist studies of open, dissipative time crystals~\cite{iemini2018boundary,booker2020non,dogra2019dissipation,kessler2021observation,taheri2022all,kongkhambut2022observation}.

Most of the theoretical and experimental work on time crystals has focused on Ising spin chain models similar to those studied in the original theoretical proposals ~\cite{Khemani2016,else2016floquet,yao2017discrete}, leaving open the question of what other types of many-body systems are capable of realizing these physics. Recent theoretical works have shown that Heisenberg spin chains can also exhibit time-crystalline behavior~\cite{PhysRevB.99.035311,li2020discrete,2021john,medenjak2020,medenjak2020scipost}, although experimental demonstrations of interaction-driven subharmonic responses in such systems have been limited to small arrays of gate-defined semiconductor spin qubits~\cite{qiao2021floquet}. Realizing substantially longer chains of highly coherent and controllable semiconductor spins will require significant technological advances which, though expected, may take some time to achieve. An alternative approach is to consider other types of many-body spin models that are realized naturally. Recently, Pal et al.~\cite{2018temporal} proposed and observed time-crystalline behavior in an NMR experiment using star-shaped molecules of various sizes, containing up to 37 spins. This system is a realization of the central-spin model, in which the satellite spins are coupled to the central spin through Ising interactions. This discovery, together with the recent results on time-crystalline behavior in Heisenberg spin chains~\cite{PhysRevB.99.035311,li2020discrete,2021john}, begs the question of whether time crystal-like phases can also exist in Heisenberg central-spin systems with either isotropic or anisotropic interactions.

This question is important for several physical systems in which central-spin systems with non-Ising interactions naturally arise. One example is color centers coupled to nuclear spin registers, which are a leading platform for quantum networks thanks to their spin-photon interfaces and long-lived nuclear spin quantum memories~\cite{Bassett2019, Bradley2019, Bourassa2020,pompili2021realization}. Here, the electronic spin at the defect site serves as the central spin, which couples to the satellite nuclear spins via anisotropic dipolar hyperfine interactions~\cite{Bradley2019,abobeih2019atomic,whaites2022adiabatic}. A second example is spins in self-assembled quantum dots, which also offer high-quality spin-photon interfaces and nuclear spin memories, making them attractive for quantum network and measurement-based quantum computing applications as well~\cite{Arakawa2020,Schwartz2016,denning2019collective,bodey2019optical,gangloff2019quantum,Michaels2021multidimensional,Gangloff2021}. Here, the central spin is a single electron or hole spin confined to the dot, while the satellite spins are surrounding nuclear spins that couple to the central electron (hole) spin via isotropic contact (anisotropic dipolar) hyperfine interactions~\cite{lee2005effect,pal2007nuclear,eble2009hole,latta2011hyperfine,urbaszek2013nuclear}. Similar types of central-spin systems are also realized in gate-defined quantum dot spin qubit platforms, although the control schemes and envisioned applications differ because such dots are not optically active; these systems are being developed predominantly as building blocks of quantum computers and simulators~\cite{Hanson2007,chekhovich2013nuclear,burkard2021semiconductor}. In all these examples, the interactions between central and satellite spins are intrinsic and unavoidable. This leads to the question of whether time crystal-like phases naturally arise in these systems when dynamical decoupling techniques based on periodic $\pi$ pulses are applied, as is commonly done to improve the coherence time of the electronic spin~\cite{Bluhm2011, Naydenov2011,taminiau2012detection, mamin2013nanoscale,Malinowski2017,Bourassa2020,zaporski2023ideal}. It is also interesting to ask whether one can one use such non-equilibrium phases to improve the performance of quantum operations, as was recently shown for quantum dot spin chains~\cite{2021john,qiao2021floquet}.

In this paper, we show that time-crystalline behavior can indeed arise in Heisenberg central-spin systems, with both isotropic and anisotropic (XXZ) interactions. Here, we define time-crystalline behavior as a subharmonic response in spin magnetizations that arises as a consequence of many-body interactions and driving, and which is robust to pulse errors and disorder. Through numerical simulations, we show that the standard Floquet pulse protocol used for Ising-coupled systems does not by itself give rise to a subharmonic response. However, we show that time-crystalline order can be induced by supplementing the Floquet driving with one of two options: either by creating a large Zeeman energy mismatch between the central and satellite spins or by applying additional pulses to the central spin every Floquet period~\cite{PhysRevB.99.035311}. Both approaches dynamically convert Heisenberg or XXZ interactions into effective Ising interactions, which can then preserve computational basis states~\cite{else2016floquet,2018temporal}. We show that pure multi-spin quantum states exhibit stable period doubling in the presence of isotropic or anisotropic interactions between the central and satellite spins when either method is used. 

The remainder of this paper is organized as follows. In Sec.~\ref{sec:ham}, we define the central-spin model Hamiltonian and discuss the parameter regimes relevant to electron-nuclear systems with hyperfine interactions. In Sec.~\ref{sec:withdis}, we study the stroboscopic dynamics of the spin expectation values using the two approaches. Firstly, we apply a large magnetic Zeeman splitting on the central spin compared to the satellite spins. Secondly, we apply additional pulses on the central spin during each Floquet period. Furthermore, we map out an effective time crystal-like phase diagram that shows when regions of stable period doubling arise as a function of interaction strength and driving errors.  We conclude in Sec.~\ref{sec:discusscon}.
\section{Model}\label{sec:ham}
\begin{figure}
    \centering
    \includegraphics[scale=0.5]{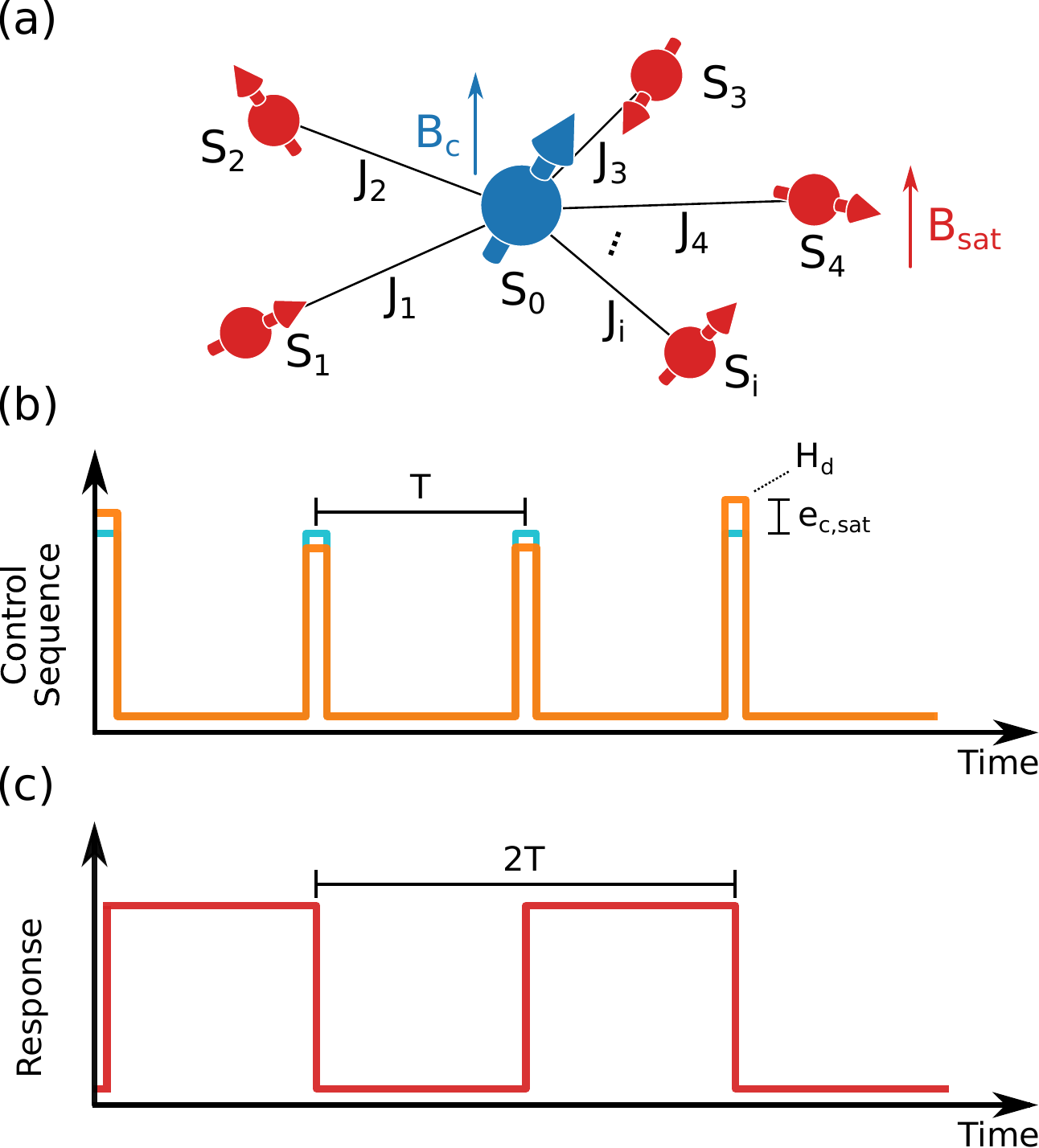}
    \caption{(a) Schematic of a central-spin system, with the central spin (blue) coupled to multiple satellite spins (red). (b) Driving sequence that is applied to each spin in the central-spin system. The driving has period $T$, and each pulse implements an imperfect $\pi$ rotation about the $x$ axis with an error in the angle equal to $e_\mathrm{c}$ for the central spin and $e_\mathrm{sat}$ for each satellite spin. (c) Schematic of the system's spin magnetization (of either the central or a satellite spin) as a function of time. Stable period doubling arises in time crystal-like phases as a consequence of periodic driving and many-body interactions despite rotation errors.}
    \label{fig:schem}
\end{figure}

We begin by defining the Hamiltonian for the XXZ-coupled central-spin model:
\begin{equation}\label{eq:hamiltonian}
 \begin{aligned}
   &H= \sum_{i=1}^{N-1}J_{xy,i}S_{x,0}S_{x,i}+\,\, \sum_{i=1}^{N-1}J_{xy,i}S_{y,0}S_{y,i}+\\
   &\,\,\sum_{i=1}^{N-1}J_{z,i}S_{z,0}S_{z,i}+B_\mathrm{c}S_{z,0}+\sum_{i=1}^{N-1}B_\mathrm{sat}S_{z,i}.\\
 \end{aligned}
\end{equation}
This model describes spin-1/2 spins such that $S_{\alpha,i}=\sigma_{\alpha,i}/2$, where $\sigma_{\alpha,i}$ is a Pauli operator ($\alpha=\{x,y,z\}$) acting on the $i$th spin. The central spin corresponds to $i=0$, while the satellite spins are labeled by $i>0$. A schematic of the model is shown in Fig.~\ref{fig:schem}. 

The central spin is coupled to each satellite spin $i$ with interaction strengths $J_{xy, i}$ and $J_{z, i}$ in the transverse and longitudinal directions, respectively, whereas the satellite spins do not interact with each other. We assume that both the transverse and longitudinal interactions $J_{xy, i}$, $J_{z, i}$ for each satellite spin $i$ are sampled from a Gaussian distribution with mean values $J_{xy}, J_z$ and variance $\delta J$. We refer to $\delta J$ as the disorder strength, and we take it to be equal for both transverse and longitudinal couplings. Throughout the paper, we use the terms isotropic or anisotropic to characterize the mean values $J_{xy}$ or $J_z$ of the couplings.

In the case of electron-nuclear central-spin systems such as NV centers in diamond coupled to surrounding ${}^{13}$C spins, the dipolar hyperfine interactions vary across nuclei because of the variation in distances between the electron and each nucleus and because of the different orientations of the displacement vector separating the two spins. Variations in hyperfine interaction strengths also arise in quantum dots because the electronic probability density can vary across nuclei. In both types of systems, the variations in interaction strengths can be modeled as disorder. $B_{\mathrm{c}}$ and $B_{\mathrm{sat}}$ are the Zeeman energies of the central and satellite spins. A difference between these energies could be due to external magnetic field gradients or due to different $g$-factors for the central and satellite spins, depending on the particular physical platform. In the main text, we neglect Zeeman splittings on satellite spins under the assumption that they are very small compared to that of the central spin; however, we examine how the subharmonic response is affected by the presence of small to moderate Zeeman splittings on satellite spins in Appendix~\ref{sec:satelliteZeeman}. There, we show that the response is non-monotonic in $B_{\mathrm{sat}}$, such that the time-crystalline behavior is either enhanced or diminished depending on its precise value.

Discrete-time crystals and related nonequilibrium phases can arise when many-body interacting systems are subject to periodic driving. Here, as in much of the previous literature~\cite{Khemani2016,else2016floquet,yao2017discrete}, we consider periodic $\pi$ pulses applied to each spin:
\begin{equation}
    H_\mathrm{d}=\sum_{k=1}^{\infty}\delta(t-kT)\left(\pi(1{-}e_{\mathrm{c}})S_{x,0}+\sum_{i=1}^{N-1}\pi(1{-}e_{\mathrm{sat}})S_{x,i}\right), \label{eq:drive}
\end{equation}
where $T$ is the driving period. We include independent pulse rotation errors, $e_{\mathrm{c}}$ and $e_{\mathrm{sat}}$, under which a time crystal-like phase should be robust~\cite{yao2017discrete}. Such errors inevitably arise from imperfect experimental control fields. The errors can in general differ between central and satellite spins since the physical mechanism used to control these spins can be distinct depending on the platform. For example, in color centers or quantum dot systems, the central electron spin and satellite nuclear spins could be driven via separate ESR and NMR control lines~\cite{morley2013quantum,pla2013high, pla2014coherent,sigillito2017all, Bradley2019, goldman2020optical,vallabhapurapu2022indirect,maity2022mechanical}. However, throughout the main text, we assume the same $\pi$ pulse driving error for both types of spins for simplicity: $e_\mathrm{c}=e_\mathrm{sat}=e_\mathrm{c,sat}$. The case of unequal pulse errors is considered in Appendix~\ref{sec:different_pulse_errors}, where the same qualitative behavior is found to emerge. In the main text, we focus on driving with instantaneous pulses as in Eq.~\eqref{eq:drive}; in Appendix \ref{sec:finite_pulses} we consider pulses of finite amplitude and duration, finding that time-crystalline behavior is still evident in this case, provided the pulse durations remain a small fraction of the driving period $T$.

We focus our study on parameter values informed by experimental implementations in quantum dots and color centers. In both cases, typical electron Zeeman splittings range from several MHz to several GHz, while hyperfine interactions range from a few hundred kHz to a few MHz. In addition to hyperfine interactions, inter-nuclear dipolar couplings are also present in these systems, with values ranging from a few Hz to a few kHz~\cite{chekhovich2015suppression,casanova2016noise}. We find that dipolar interactions (modeled as nuclear-nuclear Ising interactions) induce only small quantitative effects, so we ignore them throughout this work.

The number of nuclear spins that critically affect the central spin may vary depending on the physical system, ranging from a few 10s in color centers up to $10^5$ in optically active quantum dots. In the numerical simulations described in subsequent sections, we consider $N=6$ spins (including the central spin) unless otherwise stated. We stress that in this work, we are not concerned with demonstrating that a time crystal phase arises in the thermodynamic limit. Rather, we aim to provide evidence that residual time-crystalline effects are evident in finite-sized systems which are relevant to quantum information technologies. In Appendix~\ref{sec:numberspins}, we show that the subharmonic response becomes more stable as $N$ is increased.

In the following results, all the simulations were performed using the QuSpin Python package for exact diagonalization of quantum many-body systems~\cite{weinberg2017quspin}.

\section{Inducing time-crystalline behavior in XXZ central-spin models}\label{sec:withdis}

In this section, we present two different ways of realizing time-crystalline behavior in Heisenberg, or more generally XXZ, central-spin models. The first approach is to create a large Zeeman splitting on the central spin while applying periodic $\pi$ pulses on all the spins. In this approach, the disorder in the interaction strength between central and satellite spins is crucial for producing stable period doubling in spin magnetizations. In the second approach, we apply additional pulses to only the central spin during each driving period. These additional pulses dynamically convert the Heisenberg or XXZ interactions into effective Ising interactions, giving rise to time-crystalline behavior.

\subsection{Zeeman-mismatched time crystal}
We first show that time-crystalline behavior can be induced by a sufficiently large Zeeman energy difference between the central and satellite spins, provided there is enough disorder in the interactions. Here, we start with an initial pure state in the $z$-basis: $\ket{\Psi(0)}=\ket{\uparrow\uparrow\downarrow\uparrow\downarrow\uparrow}$. This choice is arbitrary; we have also tried other pure states in the computational ($z$) basis and observed no significant difference in the results. In Appendix~\ref{sec:initial}, we show that similar findings occur for any product state in the $z$-basis. Moreover, we show in Appendix~\ref{sec:mixed} that even if the satellite spins start out in a mixed state, a clear subharmonic response can still emerge. In this section, all of our results are averaged over 100 independent coupling disorder realizations.

The system evolves via repeated application of the Floquet operator $U_F=U_\pi U_H$, where $U_H=e^{-iHT}$ corresponds to free evolution under the central-spin model, Eq.~\eqref{eq:hamiltonian}, for interaction time $T$, and $U_\pi = \prod_i e^{-i\pi(1-e_{\mathrm{sat}}) S_{x,i}} e^{-i\pi(1-e_{\mathrm{c}}) S_{x,0}}$ is the evolution operator corresponding to a single round of pulses applied to all spins. We look for a subharmonic response in the expectation values of the components of the central and satellite spins along the magnetic field direction $z$ (i.e., spin magnetizations). To make this response more transparent, we compute these expectation values stroboscopically (i.e., after every Floquet period $T$), and we include a minus sign after every other period in anticipation of period doubling. In particular, we compute $|\langle(-1)^{n}S_{z,0}\rangle|$ for the central spin and $S_{\mathrm{avg}} = \frac{1}{N-1}\sum_{i>0}|\langle(-1)^{n}S_{z,i}\rangle|$ for the satellite spins, where the latter is averaged over all $N-1$ satellite spins, and $n$ is the number of Floquet periods.  Here, we choose to calculate the mean value of the absolute magnetization of the satellite spins for ease of presentation. We focus on absolute values of magnetization to keep the figures simple, avoiding negative values in the thermalization region. We are particularly interested in how these quantities depend on the central-spin Zeeman splitting and the average interaction strengths since the subharmonic response should only emerge when these are sufficiently large. 
\begin{figure}
       \includegraphics[width=\columnwidth]{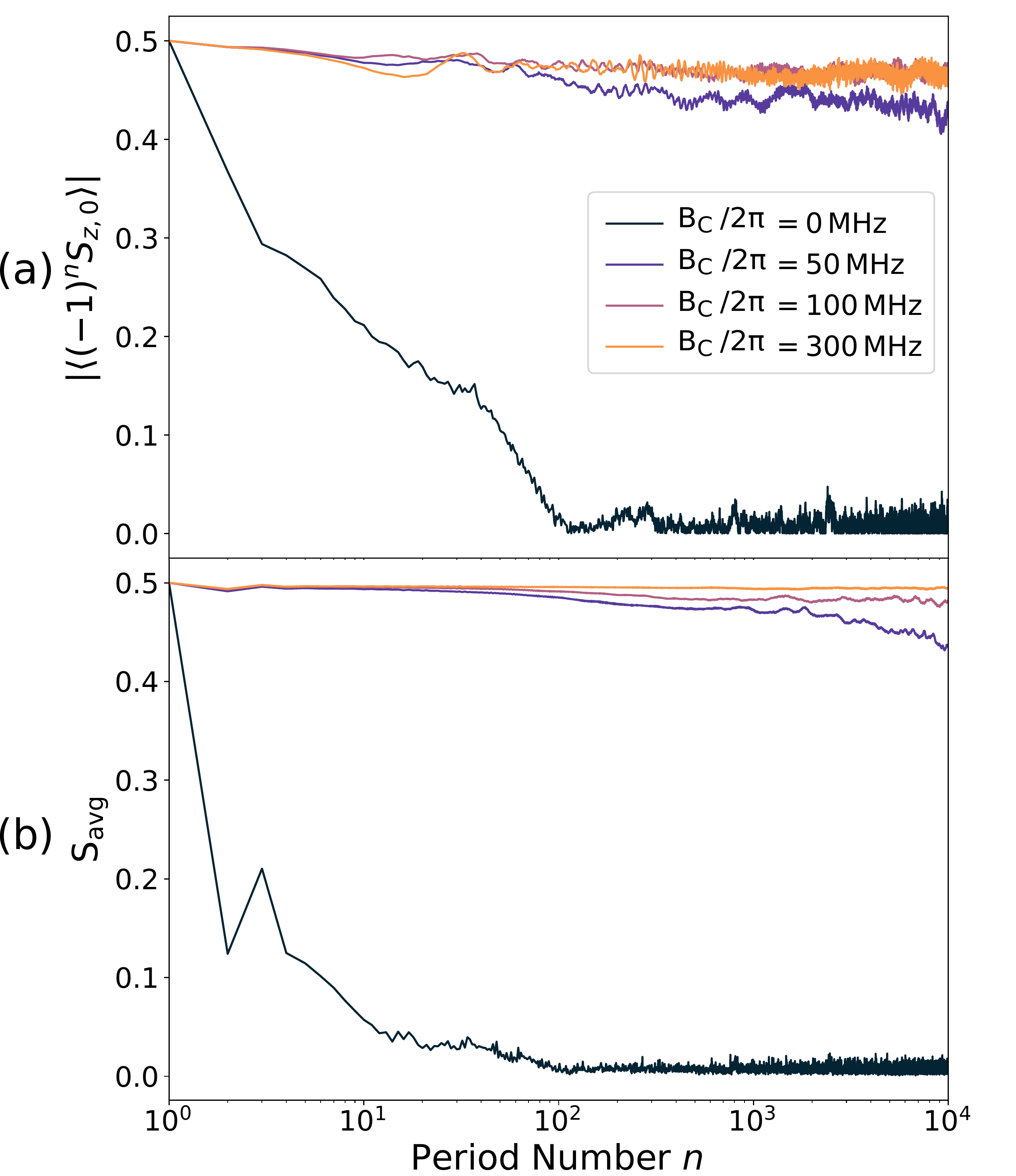}
       \caption{Emergence of period doubling with increasing central-spin Zeeman splitting $B_\mathrm{c}$ in a periodically driven central-spin model with isotropic Heisenberg interactions $J_{xy}/2\pi=J_z/2\pi=1$ MHz in the presence of $\pi$ pulse driving error $e_\mathrm{c,sat}=0.05$ and with $B_\mathrm{sat}=0$. (a) Central-spin and (b) satellite-spin magnetizations are shown. Here, $\delta J/2\pi=0.2$ MHz and $T=1~\mu$s.}
       \label{fig:isotropybc050100300}
\end{figure}  

\begin{figure*}
\includegraphics[width=\textwidth]{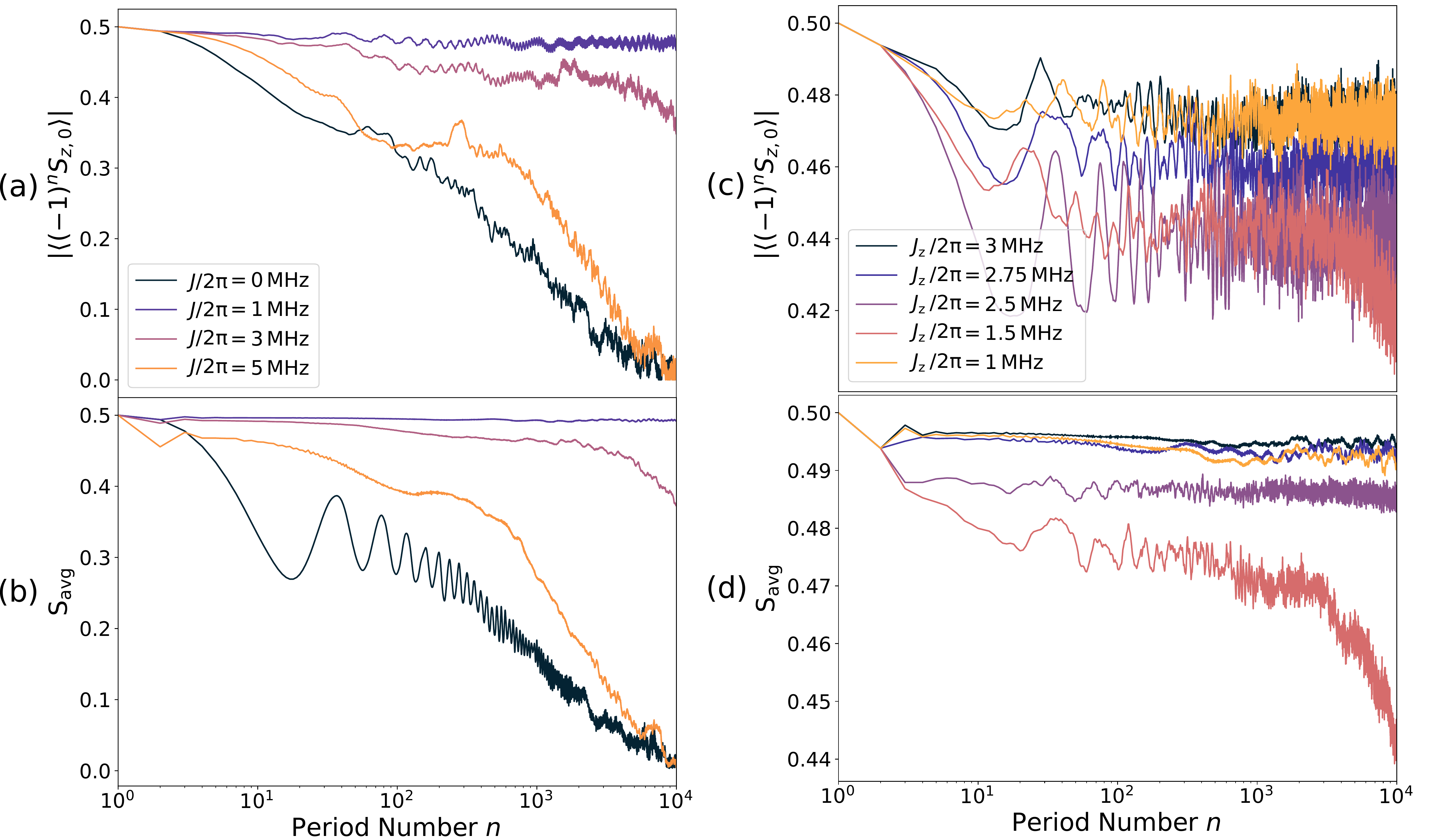}
\caption{Emergence of period doubling with increasing interaction strength in a periodically driven, Zeeman-mismatched central-spin model with (a,b) isotropic Heisenberg interactions and (c,d) anisotropic XXZ interactions. (a,b) Central- and satellite-spin magnetization as a function of the number $n$ of Floquet pulses for several values of the interaction strength $J=J_{xy}=J_z$. (c,d) Central- and satellite-spin magnetization as a function of the number $n$ of Floquet pulses for several values of $J_z$ with the total interaction strength fixed to $J/2\pi= 2J_{xy}/2\pi + J_z/2\pi=3$ MHz. The $\pi$ pulse driving error is $e_\mathrm{c,sat}=0.05$, the Zeeman energies are $B_\mathrm{c}/2\pi=300$ MHz and $B_\mathrm{sat}=0$, the disorder strength is $\delta J/2\pi=0.2$ MHz, and the driving period is $T=1~\mu$s.}
\label{fig:ZeemanDTC_varying_interactions}
\end{figure*}

We first examine how the central- and satellite-spin magnetizations depend on the strength of the central-spin Zeeman splitting $B_\mathrm{c}$, which is shown in Fig.~\ref{fig:isotropybc050100300}. We observe that as we increase the central-spin Zeeman splitting, a period-doubling effect emerges and persists out to a time scale that grows rapidly with $B_\mathrm{c}$. This indicates that applying a sufficiently strong magnetic field on the central spin is enough to induce time-crystalline behavior in Heisenberg-coupled central-spin systems.  Here, we set $B_\mathrm{sat}=0$, because, in electron-nuclear spin systems, the nuclei have $g$-factors that are orders of magnitude smaller than electronic $g$-factors. However, in systems where $B_\mathrm{sat}$ is comparable to the central-satellite spin coupling, it can still have an effect on the time-crystalline behavior. This is analyzed in Appendix~\ref{sec:satelliteZeeman}, where we show that $B_\mathrm{sat}$ can effectively enhance or diminish the longitudinal coupling, and thus modify the subharmonic response in these cases.

\begin{figure}
    \includegraphics[width=\columnwidth]{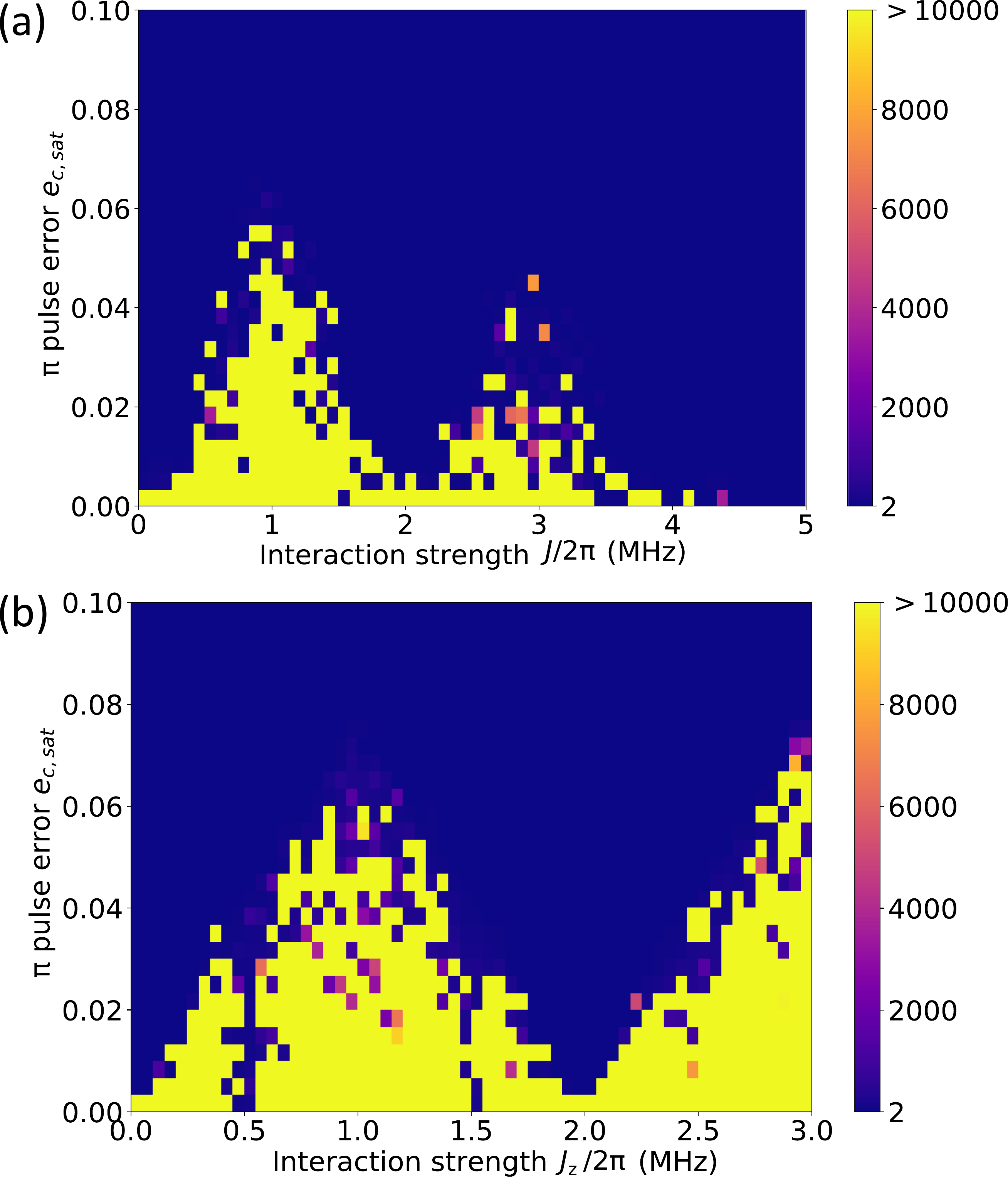}
    \caption{Phase diagrams for Zeeman-mismatched time crystals. The number of Floquet cycles (color bar) over which the return probability of the full central-spin system evolves stroboscopically (see main text for precise definition) as a function of the pulse error $e_\mathrm{c,sat}$ and (a) the interaction strength $J=J_{xy}=J_z$ of the isotropic system or (b) the longitudinal interaction strength $J_z$ of the anisotropic system with $J/2\pi=2J_{xy}/2\pi+J_z/2\pi=3$ MHz held fixed. The Zeeman energies are $B_\mathrm{c}/2\pi=300$ MHz and $B_\mathrm{sat}=0$, the disorder strength is $\delta J/2\pi=0.2$ MHz, and the driving period is $T=1~\mu$s.}
    \label{fig:numberoffloquet}
\end{figure}

A defining feature of time-crystalline physics is that the subharmonic response should only arise in the presence of sufficiently strong many-body interactions. To confirm that this is indeed the case here, we sweep the average interaction strengths $J_{xy}$, $J_z$ while keeping constant the interaction time $T=1~\mu$s. In Fig.~\ref{fig:ZeemanDTC_varying_interactions}(a,b), we observe that for isotropic interactions, the initial state is not preserved in the absence of central-satellite spin interactions as expected. However, in the parameter regime considered here, when the interactions are switched on with strength $J/2\pi=J_z/2\pi=J_{xy}/2\pi=1$ MHz, the subharmonic response in both the central- and satellite-spin magnetizations persists out to thousands of Floquet periods. Moreover, we see from the figure that as the interaction strength is increased further beyond this point, the time-crystalline behavior is destabilized, indicating that there is a finite range of interaction strengths over which a robust period doubling emerges. This can also be seen from a spectral analysis of the Floquet operator (Appendix~\ref{sec:resonances}). In the time crystal phase region, the Floquet eigenvalues come in diametrically opposite pairs~\cite{khemani2019briefhistory}. When the interaction strength is made too large or too small, the eigenvalues deviate from this simple pattern, destroying the period doubling. Below, we construct a phase diagram that delineates this region of stability (see, e.g., Fig.~\ref{fig:numberoffloquet}).

While isotropic Heisenberg interactions naturally arise in the context of electron-nuclear contact hyperfine interactions or electron-electron exchange couplings, other types of spin-spin interactions such as dipolar couplings are anistropic~\cite{Hanson2007,urbaszek2013nuclear, Bassett2019, Bradley2019}. In Fig.~\ref{fig:ZeemanDTC_varying_interactions}(c,d), we show that the temporal order is evident regardless of the amount of anisotropy. The figure shows spin expectation values for various degrees of anisotropy in the central-satellite spin couplings. In particular, we fix the total magnitude $J$ of the interactions to $J/2\pi= 2J_{xy}/2\pi + J_z/2\pi=3$ MHz, and we vary $J_{z}$ to study the effects of anisotropy. We set $B_\mathrm{c}/2\pi=300$ MHz which, as shown above, is large enough to induce time-crystalline  behavior.  We observe that for all values of $J_z$ in this range, the system exhibits a fairly stable subharmonic response. Here, we see that the greatest stability occurs in the extreme cases of purely isotropic (Heisenberg) or fully anisotropic (Ising) interactions, while for more generic types of XXZ interactions in between these extremes, the temporal order decays more rapidly. This is not generic behavior or consequence of the symmetries of the interaction, but rather occurs because of the particular parameters we have chosen. This is clarified further below, where we construct effective phase diagrams and show that two phase regions emerge around $J_zT=2\pi$ and $J_zT=6\pi$. For $T=1~\mu$s, these correspond to $J_z/2\pi=1,3$ MHz, which correspond to Heisenberg and Ising interactions, respectively, when we fix the total interaction strength to $J/2\pi=3$ MHz as we have done here. The value of $J_zT$ is what is crucial to the time-crystalline behavior, not the form of the XXZ interaction. Importantly, we can always tune the system into the centers of these phase regions by tuning the pulse period $T$, regardless of what the actual coupling strengths are in a specific system. In Appendix~\ref{sec:numberspins}, we give an approximate analysis of our quantum many-body system after two Floquet periods. We find that specific values of $J_zT$, such that the periodic driving is commensurate with the interaction strength, are needed to suppress the $\pi$ pulse driving error.

To better understand the range of interaction strengths in which temporal order arises, and to also demonstrate the stability of this order against pulse errors, we construct phase diagrams. Since we are particularly interested in the time scales over which this order persists, we define these diagrams in terms of the return probability $P(t)=|\bra{\Psi(0)}\ket{\Psi(t)}|^2$, where $\ket{\Psi(0)}$ is the initial state, and $\ket{\Psi(t)}$ is the time-evolved state of the entire system. We construct the phase diagram by counting the (even) number of Floquet cycles $n$ for which $P(2kT)\geq$ 0.95 for all $k\le n/2$ , and such that $P((2\ell+1)T)\leq$0.05 for all $\ell<n/2-1$. In simpler terms, we calculate the number of periods over which the system evolves stroboscopically to within a 5\% error. This threshold is of course arbitrary, and other values could be considered, although the results would change negligibly as will become evident from the results shown below.

Figure~\ref{fig:numberoffloquet}(a) shows the resulting phase diagram in the case of isotropic central-satellite spin interactions. It is clear from the figure that the largest degree of stability is achieved near $J/2\pi=J_{xy}/2\pi=J_z/2\pi=1$ MHz for the parameters considered, where the system can tolerate pulse errors up to nearly $e_\mathrm{c,sat}=0.06$ (i.e., 6\%). Interestingly, we also see that a second region of robust period doubling also emerges around $J/2\pi=3$ MHz, although it is not quite as insensitive to pulse errors as the first region. We also note that as either $J$ or $e_\mathrm{c,sat}$ is tuned away from these robust regions, the time scale over which the subharmonic response persists changes abruptly by orders of magnitude, from ${>}10^4$ Floquet periods down to ${<}10$ periods, indicating that these phase regions are sharply defined, despite the fact that the system consists of only $N=6$ spins. We further notice that for $J=0$, $J/2\pi=2$ MHz, or $J/2\pi\ge4$ MHz, there is virtually no robustness to pulse errors, showing that not only are many-body interactions critical to the emergence of this phenomenon, but also their precise strength. The absence of a subharmonic response at $J/2\pi$=2 and 4 MHz can be understood from the structure of the spectrum of the Floquet operator, as we show in Appendix~\ref{sec:resonances}. We can also notice that in the absence of transverse couplings and disorder, the evolution operator generated by $H$ is periodic in $J_zT$ with period $4\pi$, and so the behavior of the system for $J_z/2\pi=2$ MHz and 4 MHz should be the same as when $J_z=0$. This periodicity is approximately preserved when the transverse couplings and disorder are restored and the driving is switched on. As discussed above, it is important to stress that our choice of $T=1~\mu$s for the pulse period is arbitrary, and more generally, the centers of the phase regions are located at $J_zT=2\pi$ and $6\pi$. This in turn allows us to tune the driving period into ``resonance" with the many-body interactions to induce a subharmonic response for any value of $J_z$. In Appendix~\ref{sec:different_pulse_errors}, we show that the same qualitative features emerge when the pulse errors are unequal, $e_\mathrm{sat}\ne e_\mathrm{c}$, albeit with small quantitative differences.

\begin{figure}
       \includegraphics[width=\columnwidth]{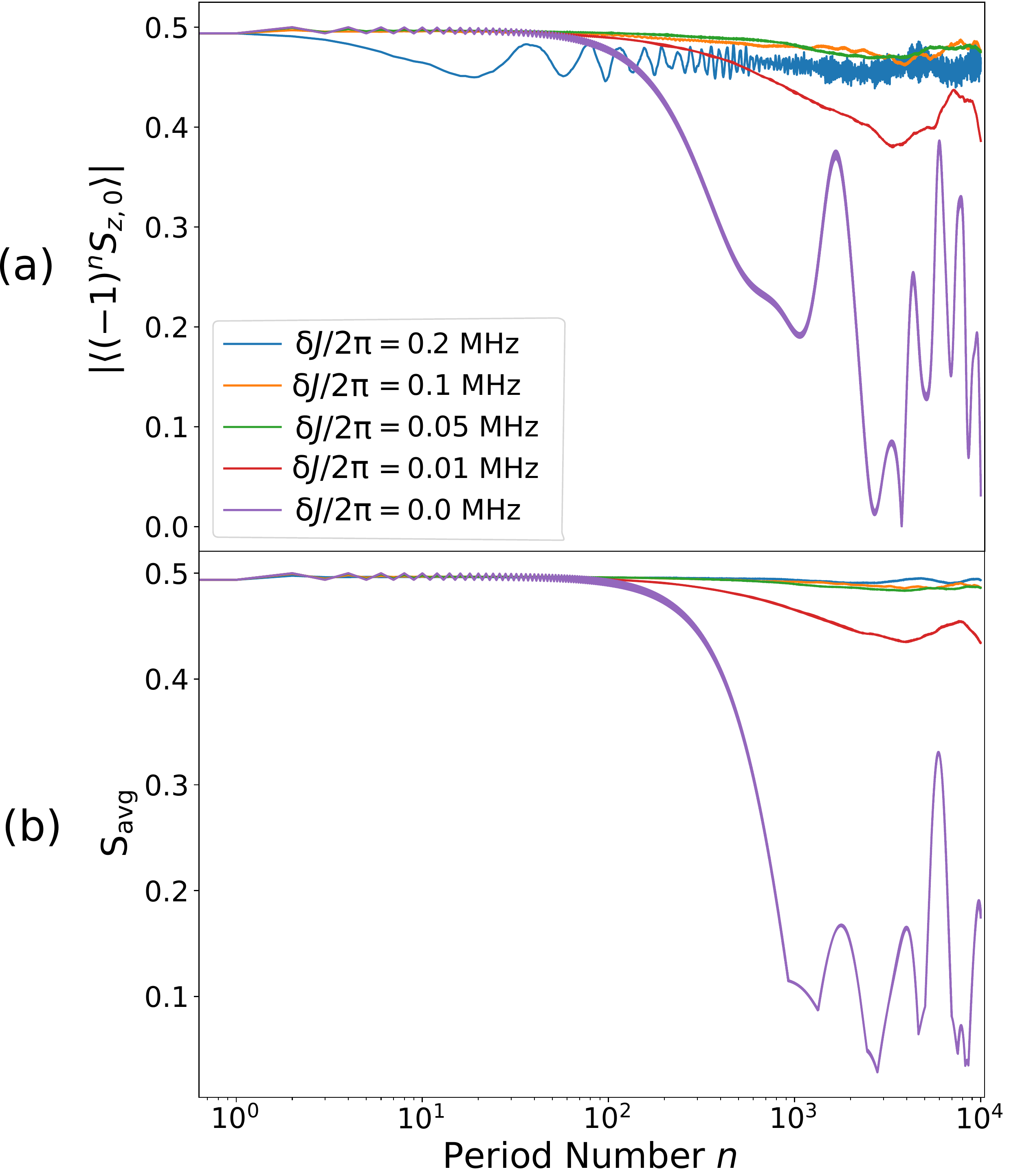}
       \caption{Role of coupling disorder in the emergence of time crystalline behavior in a central-spin model with isotropic Heisenberg interactions $J_{xy}/2\pi=J_z/2\pi=1$ MHz. (a) Central-spin and (b) satellite-spin magnetizations are shown as a function of the number of Floquet periods $n$ for several different values of the disorder strength $\delta J$. The $\pi$ pulse driving error is $e_\mathrm{c,sat}=0.05$, the Zeeman energies are $B_\mathrm{c}/2\pi=300$ MHz and $B_\mathrm{sat}=0$, and the driving period is $T=1~\mu$s.}
       \label{fig:disorderscale}
\end{figure}    

\begin{figure*}
\includegraphics[width=\textwidth]{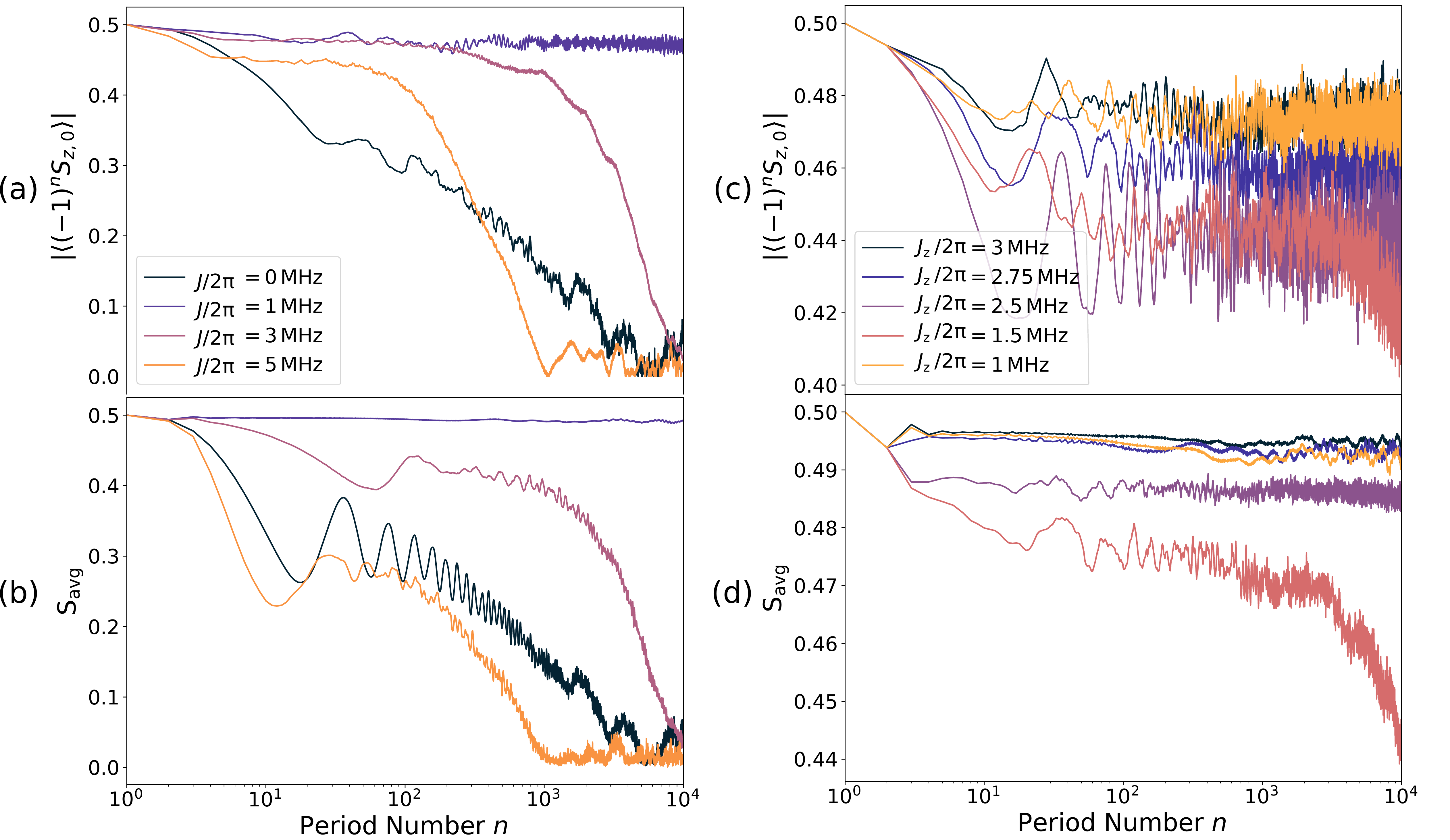}
\caption{Emergence of period doubling with increasing interaction strength in an H2I-driven central-spin model with (a,b) isotropic Heisenberg interactions and (c,d) anisotropic XXZ interactions. (a,b) Central- and satellite-spin magnetization as a function of the number $n$ of Floquet pulses for several values of the interaction strength $J=J_{xy}=J_z$. (c,d) Central- and satellite-spin magnetization as a function of the number $n$ of Floquet pulses for several values of $J_z$ with the total interaction strength fixed to $J/2\pi= 2J_{xy}/2\pi + J_z/2\pi=3$ MHz. In all cases, 40 H2I pulses per Floquet period are applied to the central spin. The Floquet driving error is $e_\mathrm{c,sat}=0.05$, the Zeeman energies are $B_\mathrm{c}=B_\mathrm{sat}=0$, the disorder strength is $\delta J/2\pi=0.2$ MHz, and the driving period is $T=1~\mu$s.}\label{fig:H2I_varying_couplings}
\end{figure*}

Figure~\ref{fig:numberoffloquet}(b) shows a phase diagram in which the one axis is the degree of coupling anisotropy rather than the total interaction strength. More specifically, we now calculate the number of stroboscopic cycles of the return probability as a function of $J_z$, with $J/2\pi=2J_{xy}/2\pi+J_z/2\pi=3$ MHz held fixed. As $J_z/2\pi$ sweeps from 0 to 3 MHz, the form of the coupling varies from an XY model with purely transversal interactions to an Ising model with only longitudinal interactions. The largest robust phase region now occurs at $J_z/2\pi=3$ MHz, corresponding to the Ising system, with insensitivity to pulse errors up to $e_\mathrm{c,sat}=0.07$ or more. A second phase region around $J/2\pi=1$ MHz is also evident which corresponds to isotropic Heisenberg interactions. This confirms what was evident from Fig.~\ref{fig:ZeemanDTC_varying_interactions}, namely that these two extreme cases exhibit the most robustness for our chosen parameters. In both cases, the time crystalline behavior extends out to more than $10^4$ Floquet periods.

In all of the above results, we assumed there is an appreciable amount of disorder in the central-satellite spin couplings ($\delta J/2\pi=0.2$ MHz). How important is this disorder to the emergence of a subharmonic response? This is addressed in Fig.~\ref{fig:disorderscale}, which shows the spin magnetizations as a function of the number of Floquet periods for amounts of disorder ranging from $\delta J=0$ to $\delta J/2\pi=0.2$ MHz. We see that the disorder has a significant effect. In particular, period doubling dissipates after only $\sim100$ periods in a disorder-free system with $\delta J=0$. On the other hand, as the disorder increases, the time scale on which the subharmonic response remains stable quickly increases to ${>}10^4$ periods for $\delta J/2\pi\geq$0.05 MHz. Thus, modest levels of disorder are necessary for the time crystalline behavior to survive on long time scales. This is consistent with time-crystalline order associated with many-body localization, in which non-ergodicity is caused by the emergence of local integrals of motion~\cite{hetterich2018detection,schliemann2021many}.

\subsection{Heisenberg to Ising pulses on central spin}

\begin{figure*}
\includegraphics[width=\textwidth]{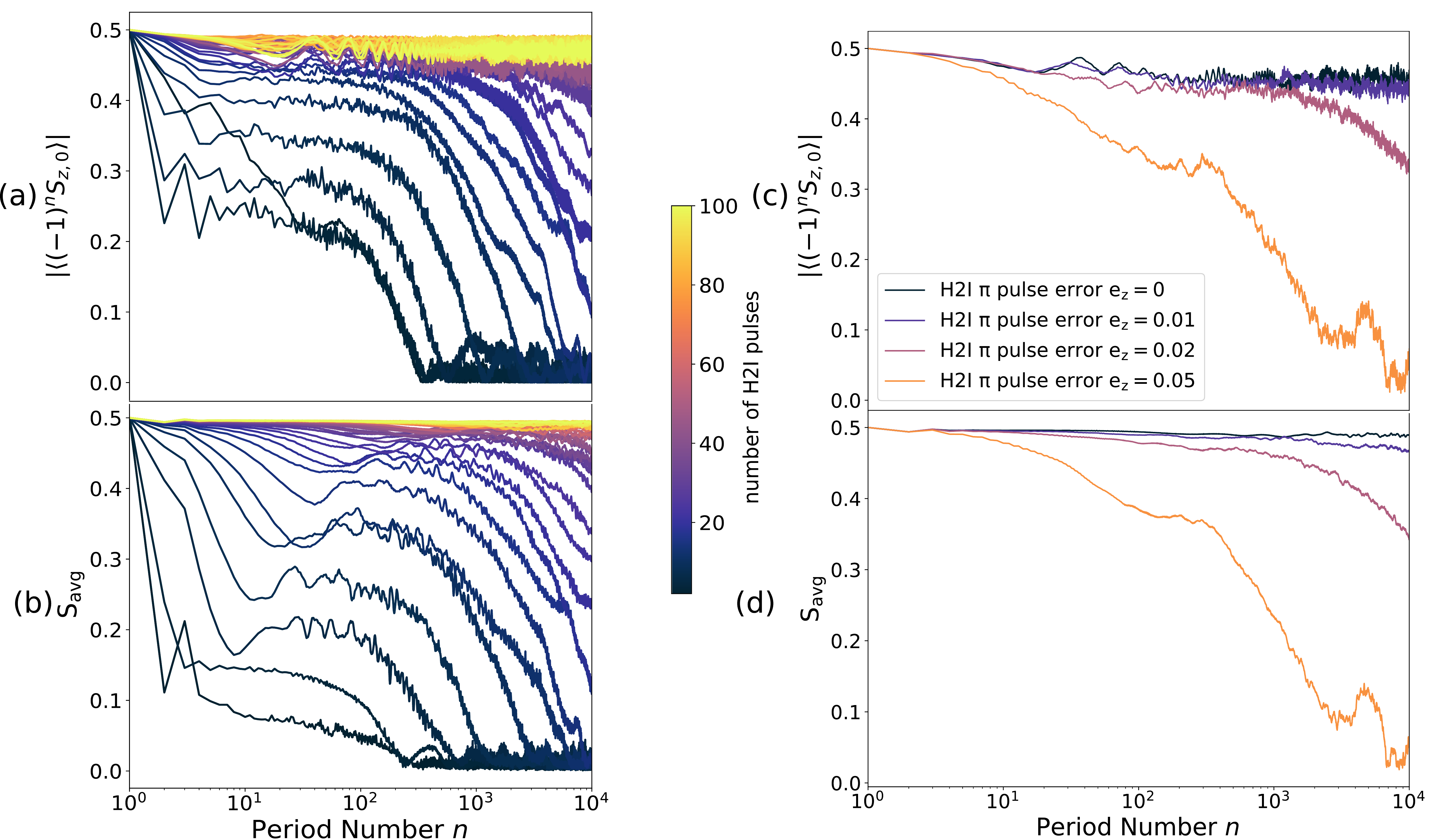}
\caption{(a,b) Emergence of period doubling in a periodically driven central-spin model in which the central spin is subject to additional H2I pulses every Floquet period. (a) Central-spin and (b) satellite-spin magnetizations are shown as a function of the number $n$ of Floquet periods for various numbers of H2I pulses ranging from 2 to 100 every Floquet period. (c,d) Stability of the H2I-induced temporal order with increasing H2I pulse error $e_\mathrm{z}$. (c) Central-spin and (d) satellite-spin magnetizations are shown as a function of the number $n$ of Floquet periods for four different values of $e_\mathrm{z}$ in the case of 40 H2I pulses per Floquet period. In all panels, the interactions are isotropic with strength $J_{xy}/2\pi=J_z/2\pi=1$ MHz, the Floquet driving error is $e_\mathrm{c,sat}=0.05$, the Zeeman energies are $B_\mathrm{c}=B_\mathrm{sat}=0$, the disorder strength is $\delta J/2\pi=0.2$ MHz, and the driving period is $T=1~\mu$s.}\label{fig:H2I_varying_numPulses_error}
\end{figure*}

In this section, we show that there is an alternative way to create a time crystal-like phase where, instead of using a large Zeeman energy mismatch between central and satellite spins, we apply additional $\pi$ pulses to the central spin every Floquet period. These additional pulses act as a dynamical decoupling sequence that dynamically suppresses two of the three interaction terms in the XXZ Hamiltonian, Eq.~\eqref{eq:hamiltonian}, resulting in an effective Ising interaction. We refer to these additional pulses as `H2I' pulses following Ref.~\cite{PhysRevB.99.035311}, which introduced a similar technique for spin chains. In the spin chain case, this echoing out of interaction terms works provided the H2I pulses are applied to every other spin, such that the pulses act on only one spin in each interacting pair. In the central-spin model, the same effect can be achieved by applying H2I pulses to only the central spin since each interacting pair of spins in this model includes the central one. Applying enough H2I pulses should then reduce the system to an effective Ising central-spin model, which was shown in Ref.~\cite{2018temporal} to exhibit time crystal-like signatures. As in the spin chain case, the larger the number of H2I pulses per Floquet period, the more the effective interaction resembles an Ising form, and the rotation axis of the H2I pulses determines the orientation of the Ising interaction. Here, we choose the H2I rotation axis to be the $z$ axis, implying that the effective Ising interaction is of $S_{z,0}S_{z,i}$ type. The Floquet operator is then given by $U_F=U_\pi U_\mathrm{H2I}(T)$, where $U_\pi$ is the same as in the previous section, while
\begin{equation}
 U_\mathrm{H2I}(T)=[e^{i\pi S_{z,0}(1-e_\mathrm{z})}U_H(T/m)]^{m},  
\end{equation}
where $U_H(t)=e^{-iHt}$ with $H$ defined in Eq.~\eqref{eq:hamiltonian}, $m$ is the number of H2I pulses, and $e_\mathrm{z}$ is the H2I rotation error. Throughout this section, we set the Zeeman energies to zero, $B_\mathrm{c}=0$, $B_\mathrm{sat}=0$, since they are no longer needed to induce temporal order. We continue to sample the couplings from Gaussian distributions with means $J_{xy}$, $J_z$ and standard deviations $\delta J$.

To confirm that the H2I technique can generate time-crystalline behavior in the central-spin model, we first compute the spin magnetizations as a function of the number of Floquet periods using 40 H2I pulses per period. The results for isotropic interactions are shown in Fig.~\ref{fig:H2I_varying_couplings}(a,b). Here, we initialize the system in the computational basis pure state  $\ket{\Psi (0)}=\ket{\uparrow\uparrow\downarrow\uparrow\downarrow\uparrow}$, which has no underlying symmetry, thus avoiding any sort of fine-tuning or bias in the results. We see from the figure that as we increase the interaction strength, a subharmonic response gradually emerges. As long as the interaction strength is sufficiently close to $J/2\pi=1$ MHz, this subharmonic response is long-lived, similarly to what we saw in the case of the Zeeman-mismatch-induced temporal order (c.f., Fig.~\ref{fig:ZeemanDTC_varying_interactions}). As discussed in the previous section, the subharmonic response is generally most stable when $J_zT=2\pi$, allowing one to tune the system into this regime for any $J_z$ by adjusting the pulse period $T$ appropriately. However, unlike in the Zeeman-mismatched case, here we do not see a revival near $J/2\pi=3$ MHz, suggesting the absence of a second region of robustness in the corresponding phase diagram. Below, we confirm that this is indeed the case. In Fig.~\ref{fig:H2I_varying_couplings}(c,d), we examine the effect of interaction anisotropy by tuning the interactions from Ising to Heisenberg form. We again find that these two extremal cases exhibit the most robustness, although Ising interactions are clearly more effective in achieving a long-lived period doubling. We also see from the figure that the stability is much weaker for generic XXZ interactions compared to the Zeeman-mismatch-induced phase.

Because the effective many-body interactions only converge to Ising form in the limit of infinitely many H2I pulses, it is important to investigate how the temporal order depends on the number of pulses. This is also an important experimental consideration since there is a limit to how many pulses can be applied in the laboratory. Figure~\ref{fig:H2I_varying_numPulses_error}(a,b) shows that as we apply more and more H2I pulses to the central spin, the subharmonic response is preserved for increasingly longer times. We see that for 40 H2I pulses per period the temporal order survives for 100s of Floquet periods, while for ${>}80$ pulses, this time scale increases by an order of magnitude or more. In Appendix~\ref{sec:h2i_pulses}, we provide some analytical intuition behind the H2I mechanism, and we show that the time scale on which the central spin magnetization is stabilized increases superlinearly with the number of H2I pulses. We also see from the figure that the satellite spins stabilize much more quickly compared to the central spin. Note that for $T=1~\mu$s (the Floquet period considered here), 100 H2I pulses correspond to a pulse spacing of 10 ns, which while experimentally feasible, likely approaches the limits of current arbitrary waveform generators. Another important experimental consideration is the role of errors in the H2I pulses. This is investigated in Fig.~\ref{fig:H2I_varying_numPulses_error}(c,d), which shows the central- and satellite spin magnetizations for errors ranging from $e_\mathrm{z}=0$ up to 0.05. We see that while errors at the level of 1\% or less ($e_\mathrm{z}\le0.01$) do not have a significant effect, larger errors quickly destroy the temporal order. Thus, the H2I pulses must be accurate to within 1\% to be effective at inducing time crystalline behavior.

\begin{figure}
\includegraphics[width=\columnwidth]{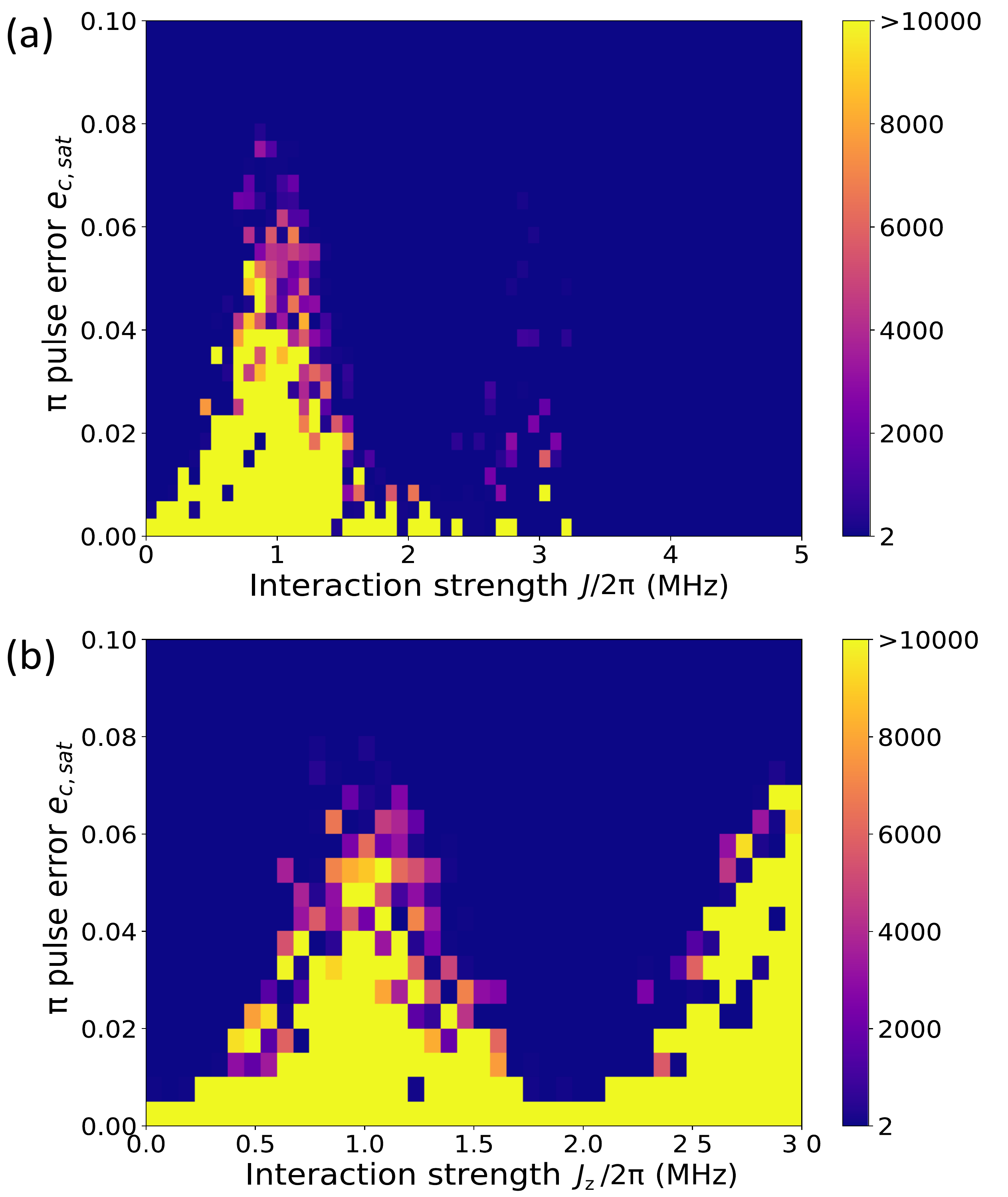}
\caption{Phase diagrams for H2I-induced time crystals with $N=4$ spins. The number of Floquet cycles (color bar) over which the return probability of the full central-spin system evolves stroboscopically (see main text for precise definition) as a function of the pulse error $e_\mathrm{c,sat}$ and (a) the interaction strength $J=J_{xy}=J_z$ of the isotropic system or (b) the longitudinal interaction strength $J_z$ of the anisotropic system with $J/2\pi=2J_{xy}/2\pi+J_z/2\pi=3$ MHz held fixed. In both panels, the central spin is subject to 80 H2I pulses per Floquet period, the Zeeman energies are $B_\mathrm{c}=B_\mathrm{sat}=0$, the disorder strength is $\delta J/2\pi=0.2$ MHz, and the driving period is $T=1~\mu$s.}\label{fig:H2I_phase_diagram}
\end{figure}

Next, we turn to constructing a phase diagram for the H2I-driven central-spin system. As in the Zeeman-mismatched case above, we define the phase diagram by counting the number of Floquet cycles over which the return probability $P$ exhibits $2T$ periodicity to within 5\% accuracy ($P\ge0.95$ after every second period). Due to computational costs, here we restrict attention to $N=4$ spins. We initialize the system in a $z$-basis pure state $\ket{\Psi(0)}=\ket{\uparrow\uparrow\downarrow\uparrow}$ and apply Floquet pulses to all spins with period $T=1~\mu$s, interspersed with 80 H2I pulses applied to only the central spin. Figure~\ref{fig:H2I_phase_diagram}(a) shows the resulting phase diagram in the case of isotropic interactions, where it is evident that a phase region centered around $J/2\pi=1$ MHz emerges, in which the time crystalline behavior is preserved out to $10^4$ Floquet periods or more. As in the case of the Zeeman-mismatched system (c.f., Fig.~\ref{fig:numberoffloquet}), this temporal order persists up to Floquet pulse errors of order $e_\mathrm{c,sat}\sim0.06$. On the other hand, the second phase region near $J/2\pi=3$ MHz is no longer evident in the H2I case. In Fig.~\ref{fig:H2I_phase_diagram}(b), we present a different phase diagram that shows how the robustness of the temporal order depends on the coupling anisotropy. The results are similar to what we found for the Zeeman-mismatched case above in Fig.~\ref{fig:numberoffloquet}, namely the temporal order is most robust near $J_z/2\pi=1$ MHz and 3 MHz, corresponding to purely Heisenberg or Ising interactions, while it quickly dissipates away from these values when the Floquet pulse errors exceed 0.5\% ($e_\mathrm{c,sat}\ge0.005$).

\section{Conclusions}\label{sec:discusscon}
In conclusion, we showed that time-crystalline behavior can arise in periodically driven central-spin models with any type of XXZ interactions. We found that, unlike in the case of Ising interactions, simple periodic driving and many-body interactions alone are insufficient to realize his behavior. For general XXZ interactions, we showed two ways to induce a stable subharmonic response in spin magnetizations: (i) creating a large Zeeman energy mismatch between central and satellite spins, or (ii) applying additional $\pi$ pulses every period to only the central spin. We found that both approaches lead to stable period doubling that survives for thousands of Floquet periods, provided the interaction strength (or equivalently the pulse period) and disorder are tuned appropriately. We found that the greatest stability arises when the pulse period is given by the inverse of the interaction strength. Our results are of direct relevance to systems in which a central electronic spin couples to surrounding nuclear spins via hyperfine interactions, as occurs in color centers or semiconductor quantum dots.

\acknowledgements
We would like to thank Bikun Li for helpful discussions. This work was supported in part by DARPA (grant no. D18AC00025). S.E.E. and M. A. acknowledge support from the EU Horizon 2020 programme (GA 862035 QLUSTER). E.B. also acknowledges support from NSF grant no. 1847078. D.A.G. acknowledges a Royal Society University Research Fellowship.

\appendix

\section{Effect of satellite-spin Zeeman splittings}\label{sec:satelliteZeeman}

\begin{figure}
\includegraphics[width=\columnwidth]{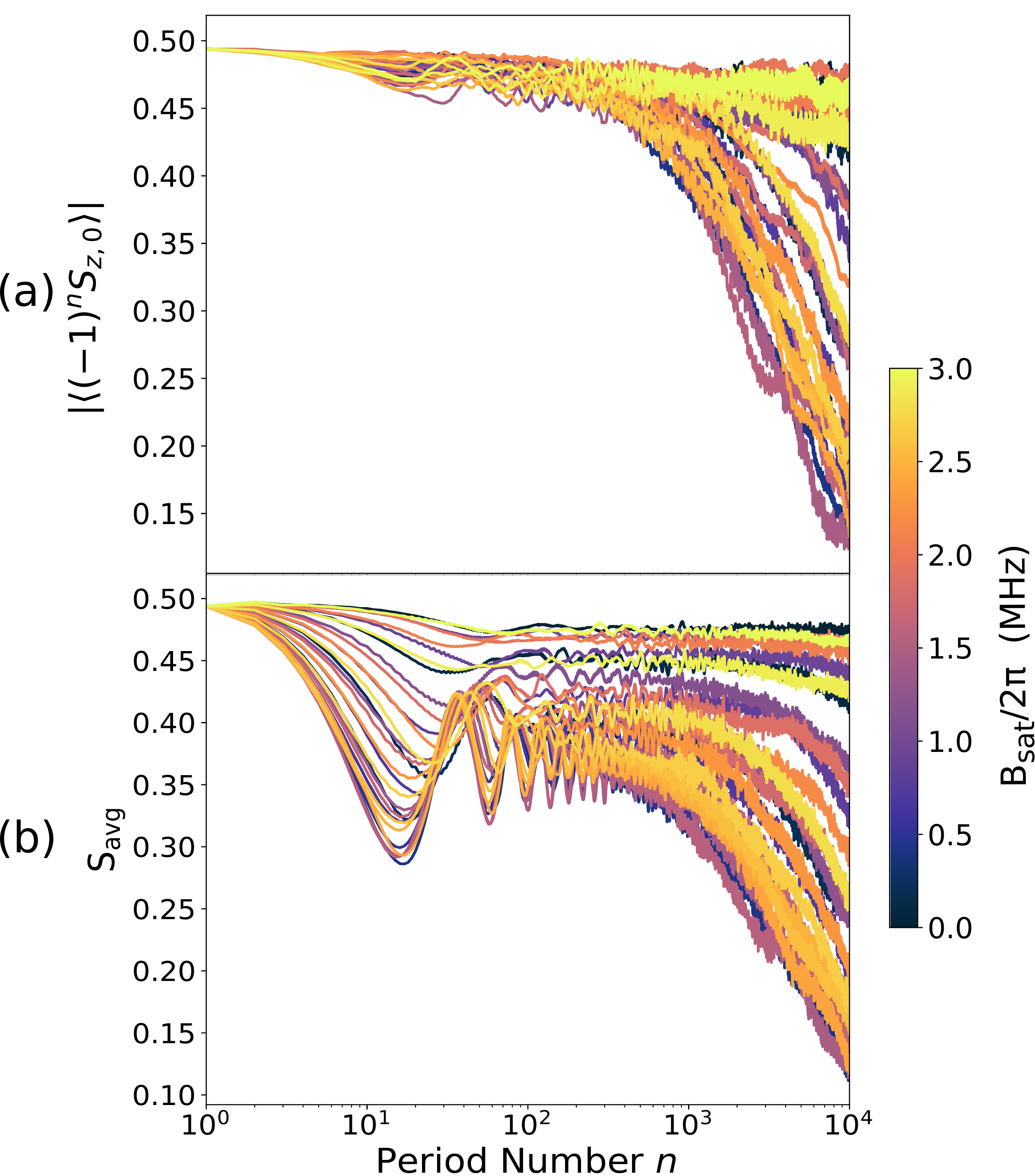}
\caption{Effect of satellite-spin Zeeman energy on time-crystalline order in a periodically driven central-spin system with isotropic interactions. Central-spin and satellite-spin magnetizations are shown as a function of the number $n$ of Floquet periods for several different values of the satellite-spin Zeeman energy $B_\mathrm{sat}/2\pi$ ranging from 0 to 3 MHz. Here, the interaction strength is $J_{xy}/2\pi=J_z/2\pi=1$ MHz, the Floquet driving error is $e_\mathrm{c,sat}$=0.05, the central-spin Zeeeman energy is $B_\mathrm{c}/2\pi=300$ MHz, the disorder strength is $\delta J/2\pi=0.2$ MHz, satellite magnetic Zeeman splitting disorder strength $\delta B_\mathrm{sat}/2\pi=0.05$ MHz, and the driving period is $T=1~\mu$s. 
}\label{fig:effect_of_Bsat}
\end{figure}

In this appendix, we examine the role of satellite-spin Zeeman splittings in the stability of the emergent time-crystalline order. To do this, we bring the system into a parameter regime in which time-crystalline behavior is evident by setting $B_\mathrm{c}/2\pi=300$ MHz and then study how this behavior changes as we increase the satellite-spin Zeeman splitting. The results are shown in Fig.~\ref{fig:effect_of_Bsat}, where it is evident that the robustness of the time-crystalline behavior is non-monotonic as a function of $B_\mathrm{sat}$. In the case of isotropic interactions, we observe time-crystal-like behavior for specific values of $B_\mathrm{sat}$. For example, in the case where $B_\mathrm{sat}/2\pi$ is an integer (0,1,2,3 MHz), a strong subharmonic response is evident. We also observe similar behavior in the case of anisotropic Heisenberg interactions with $J_{z}/2\pi=3$ MHz, $J_{xy}/2\pi=1$ MHz. However, for other values of $B_\mathrm{sat}$, the stroboscopic dynamics decays much more quickly. This can be understood as follows. Due to the high on-site magnetic field, we can neglect electron-nuclear flip-flop terms and approximate our Hamiltonian as
\begin{equation}\label{eq:hamiltonianIsingbsat}
 \begin{aligned}
   &H\approx J_{z,i}\,\,S_{z,0}\sum_{i=1}^{N-1}S_{z,i}+B_\mathrm{c}S_{z,0}+B_\mathrm{sat}\sum_{i=1}^{N-1}S_{z,i}.\\
 \end{aligned}
\end{equation}
We observe from this approximate Hamiltonian that the inclusion of satellite-spin Zeeman splittings can effectively enhance or diminish the longitudinal coupling depending on the state of the central spin. Specifically, the effective longitudinal coupling is
\begin{equation}
 \begin{aligned}
     & J_{z,i}^{\mathrm{eff}}=J_{z,i}+2 B_\mathrm{sat} \rm{\,\,if\,\, central\,\, spin\,\, is\,\, in\,\, \ket{\uparrow}\,\, state}\\
     &J_{z,i}^{\mathrm{eff}}=J_{z,i}-2B_\mathrm{sat}\rm{\,\,if\,\, central\,\, spin\,\, is\,\, in\,\, \ket{\downarrow}\,\, state}\\
 \end{aligned}    
\end{equation}
Thus, for initial states in which the central spin is $\ket{\uparrow}$, as $B_\mathrm{sat}$ increases, the effective coupling increases, bringing the system into and out of time crystal-like phase regions. This is why we see a subharmonic response in the presence of specific values of $B_\mathrm{sat}$ in Fig.~\ref{fig:effect_of_Bsat}.

{

\section{Effect of differing pulse errors on central and satellite spins}\label{sec:different_pulse_errors}

\begin{figure}
    \includegraphics[scale=0.285]{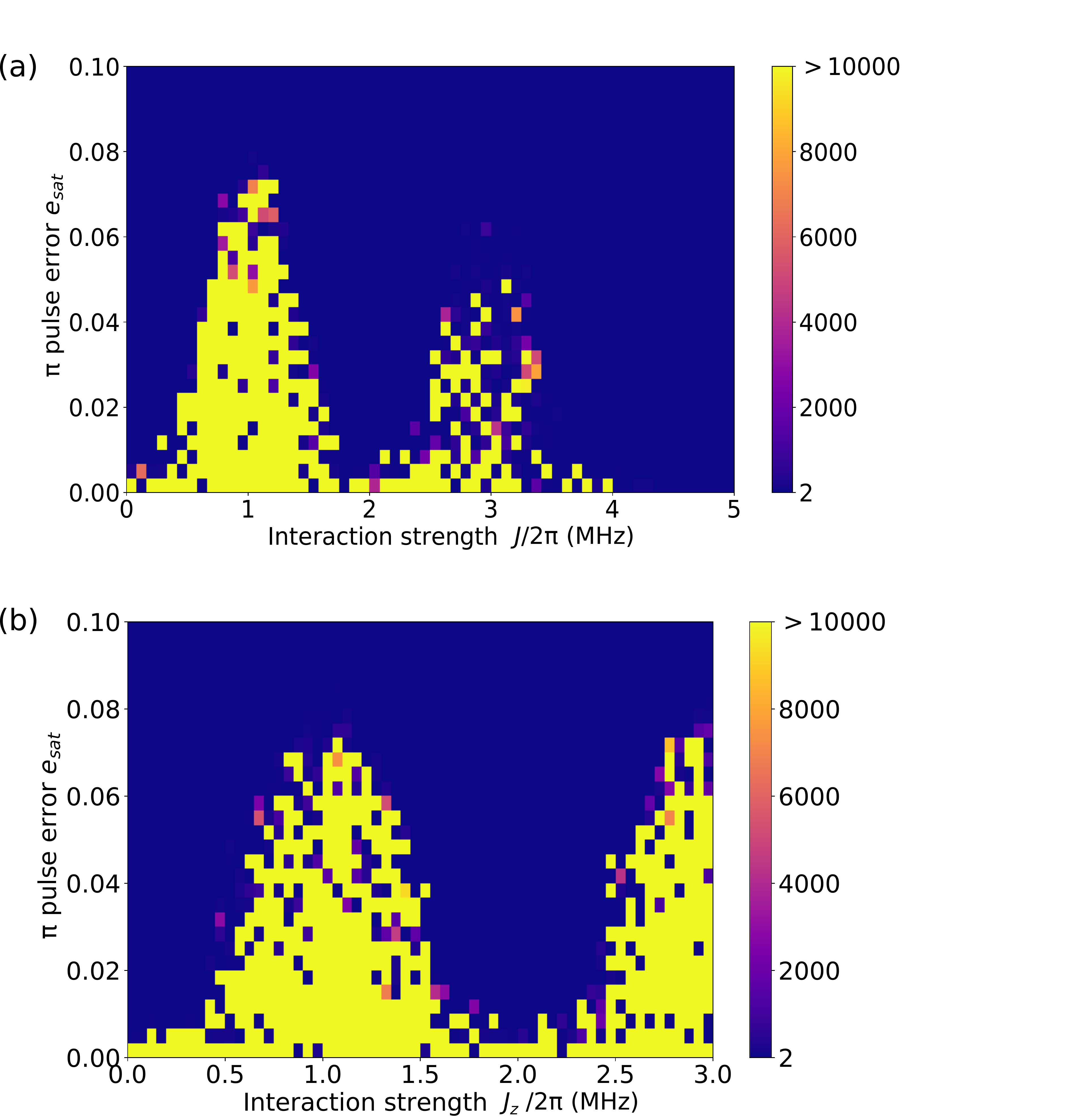}
    \caption{Phase diagrams for Zeeman-mismatched time crystals with fixed central spin pulse error. The number of Floquet cycles (color bar) over which the return probability of the full central-spin system evolves stroboscopically (see main text for precise definition) as a function of the pulse error $e_\mathrm{sat}$ with a fixed central spin pulse error $e_\mathrm{c}=0.01$ and (a) the interaction strength $J=J_{xy}=J_z$ of the isotropic system or (b) the longitudinal interaction strength $J_z$ of the anisotropic system with $J/2\pi=2J_{xy}/2\pi+J_z/2\pi=3$ MHz held fixed. The Zeeman energies are $B_\mathrm{c}/2\pi=300$ MHz and $B_\mathrm{sat}=0$, the disorder strength is $\delta J/2\pi=0.2$ MHz, and the driving period is $T=1~\mu$s.}
    \label{fig:numberoffloquetec001}
\end{figure} 

In experimental realizations, the $\pi$ pulse errors may differ between the central and satellite spins. To investigate the impact of such differences, we consider a phase diagram for isotropic Heisenberg interactions in which only $e_\mathrm{sat}$ is allowed to vary, while $e_\mathrm{c}$ remains fixed. The result is shown in Fig.~\ref{fig:numberoffloquetec001}(a), where the pulse error on the central spin is held constant at $e_\mathrm{c}=0.01$. This value is consistent with recent demonstrations of single-qubit gates in central spin qubits~\cite{yoneda2018quantum,Bradley2019,cerfontaine2020closed} and satellite spin qubits~\cite{chekhovich2020nuclear,hensen2020silicon}. As in the case of equal pulse errors, we see a subharmonic response persist over a large number of Floquet periods for interaction strengths near $J/2\pi =1, 3$ MHz. However, in this case, due to the relatively small central spin error $e_\mathrm{c}=0.01$, the subharmonic response lasts for $10^4$ Floquet periods up to satellite pulse errors of up to $e_\mathrm{sat}=0.08$. We further notice again that for $J=0$, $J/2\pi=2$ MHz, or $J/2\pi\ge4$ MHz, there is no preservation of the initial state, and so we have the same dependence on the interaction strength as we found for equal pulse errors. In the case of anisotropic Heisenberg interactions, we again find that the phase diagram bears a qualitative resemblance to the equal-error case, as shown in Fig.~\ref{fig:numberoffloquetec001}(b).

\begin{figure}
\includegraphics[scale=0.285]{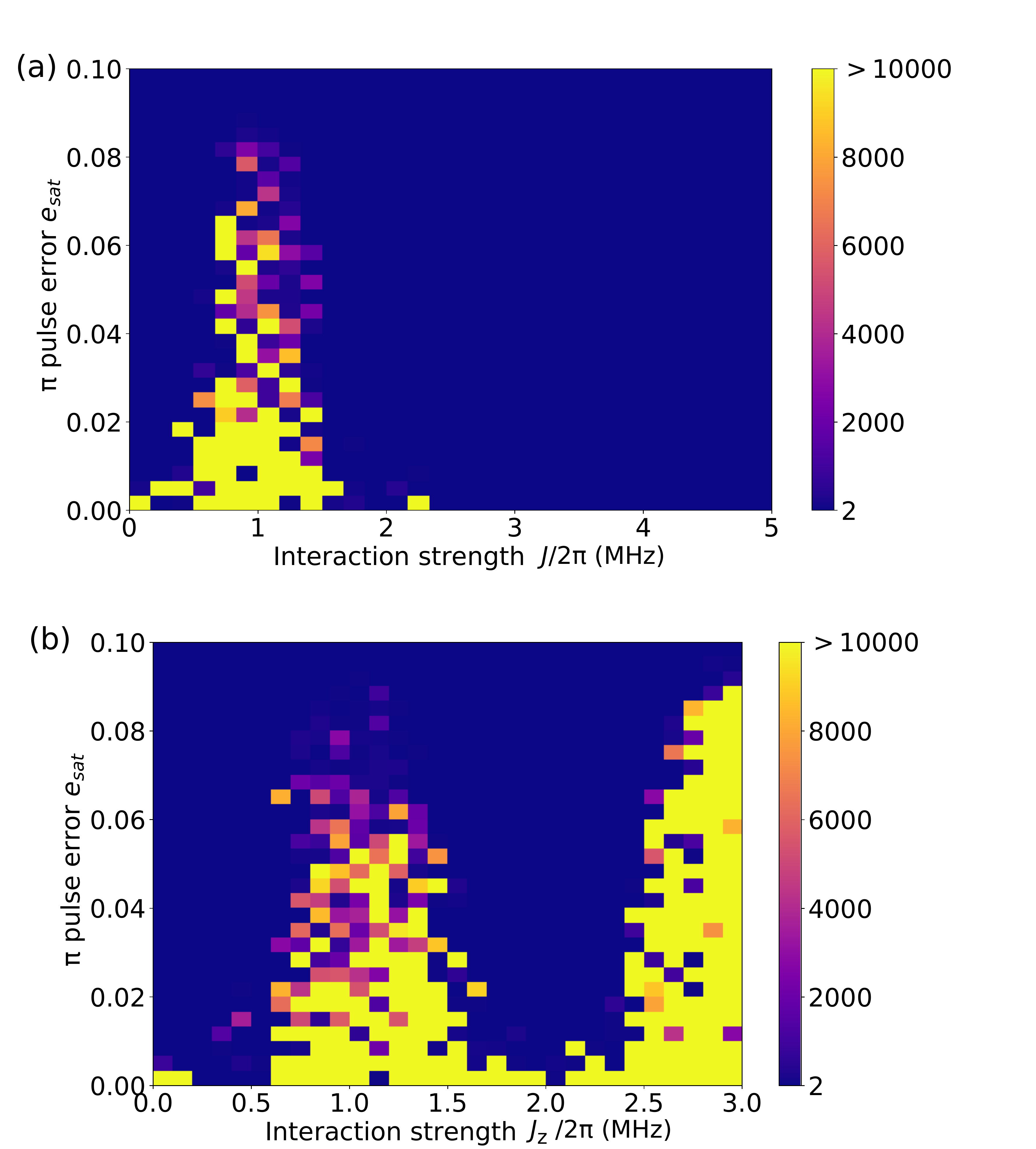}
\caption{Phase diagrams for H2I-induced time crystals with fixed central spin pulse error for $N=4$ spins. The number of Floquet cycles (color bar) over which the return probability of the full central-spin system evolves stroboscopically (see main text for precise definition) as a function of the pulse error $e_\mathrm{sat}$ with a fixed central spin pulse error $e_\mathrm{c}=e_\mathrm{z}=0.01$ and (a) the interaction strength $J=J_{xy}=J_z$ of the isotropic system or (b) the longitudinal interaction strength $J_z$ of the anisotropic system with $J/2\pi=2J_{xy}/2\pi+J_z/2\pi=3$ MHz held fixed. In both panels, the central spin is subject to 60 H2I pulses per Floquet period, the Zeeman energies are $B_\mathrm{c}=B_\mathrm{sat}=0$, the disorder strength is $\delta J/2\pi=0.2$ MHz, and the driving period is $T=1~\mu$s.}\label{fig:H2I_phase_diagramec001}
\end{figure}

We can also consider the impact of unequal pulse errors on time crystals induced by H2I driving. Specifically, we map out the phase diagram for isotropic Heisenberg interactions in Fig.~\ref{fig:H2I_phase_diagramec001}(a), finding similar behavior as in the previous Fig.~\ref{fig:numberoffloquetec001}(a), even with errors on both the Floquet and H2I pulses. The subharmonic response is preserved for more than $10^4$ Floquet periods, especially in the vicinity of $J/2\pi=1,3$ MHz.

In the case of anisotropic Heisenberg interactions, we sweep the $J_z$ interaction strength with the total interaction strength fixed to $J/2\pi= 2J_{xy}/2\pi + J_z/2\pi=3$ MHz, mapping out a phase diagram for a fixed central spin error $e_\mathrm{c}=e_\mathrm{z}=0.01.$ In Fig.~\ref{fig:H2I_phase_diagramec001}(b) we observe similar results as in the case of the large Zeeman-mismatched time crystal, with the most robust region in the case of pure Heisenberg or pure Ising interactions up to $8\%$ $\pi$ pulse driving error on the satellite spins.

\section{$\pi$ pulses with finite duration and amplitude}\label{sec:finite_pulses}

\begin{figure}[tbph]
    \centering
    \includegraphics[scale=0.3]{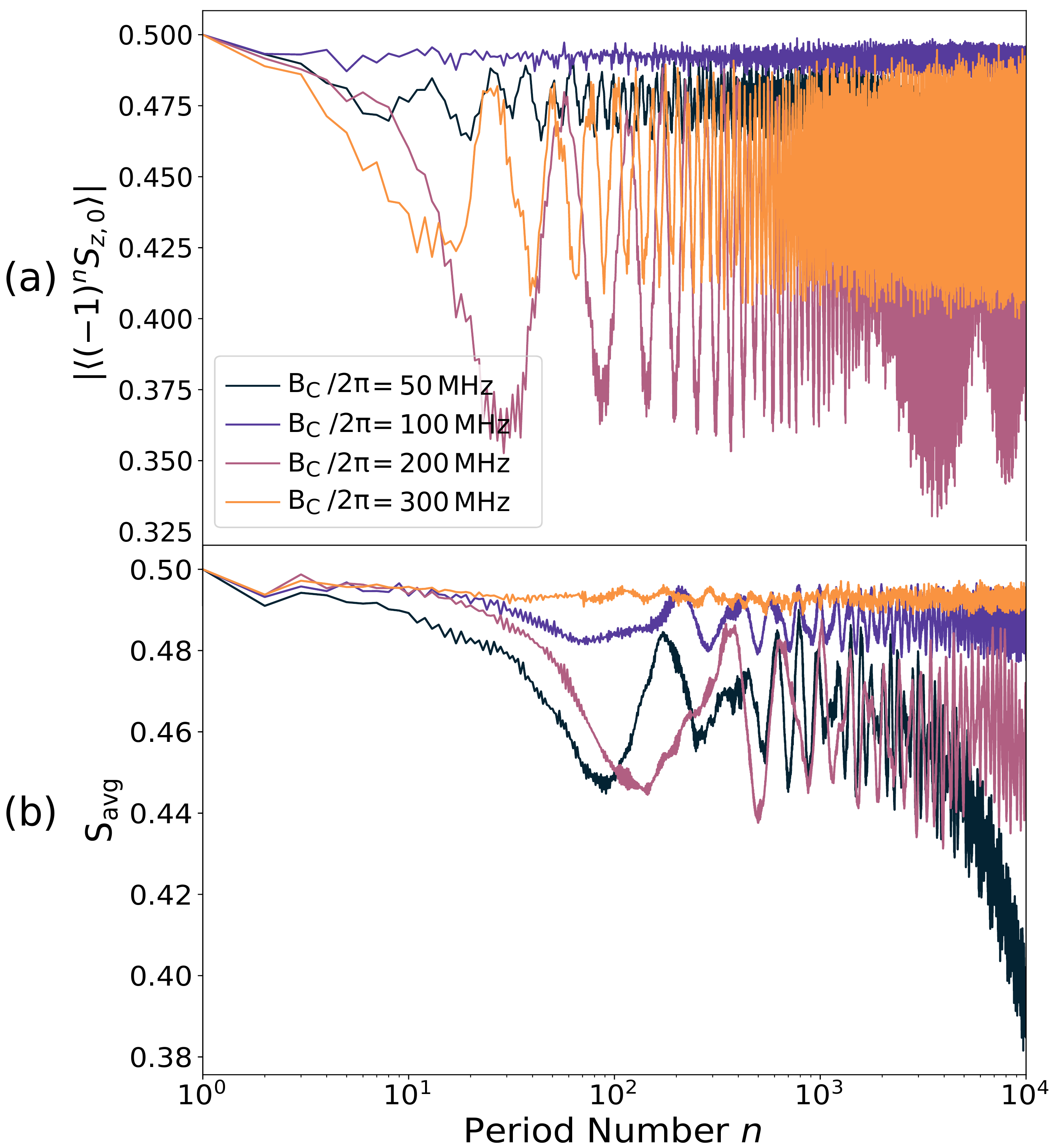}
    \caption{Effect of using finite-amplitude pulses as in Eq.~\eqref{eq:sxnon}. (a) Central-spin and (b) satellite-spin magnetizations for a time-dependent periodically driven central-spin model with isotropic interactions.  We consider isotropic Heisenberg interactions $J_{x,y}/2\pi=J_z/2\pi=J/2\pi=1$ MHz, the Floquet driving error is $e_\mathrm{c,sat}=0.05$, and the driving period is $T=1~\mu$s. The pulse time is $\eta T$= 0.1 ns.}
    \label{fig:pulsedifferentfield}
\end{figure}
\begin{figure}[tbph]
    \centering
    \includegraphics[scale=0.32]{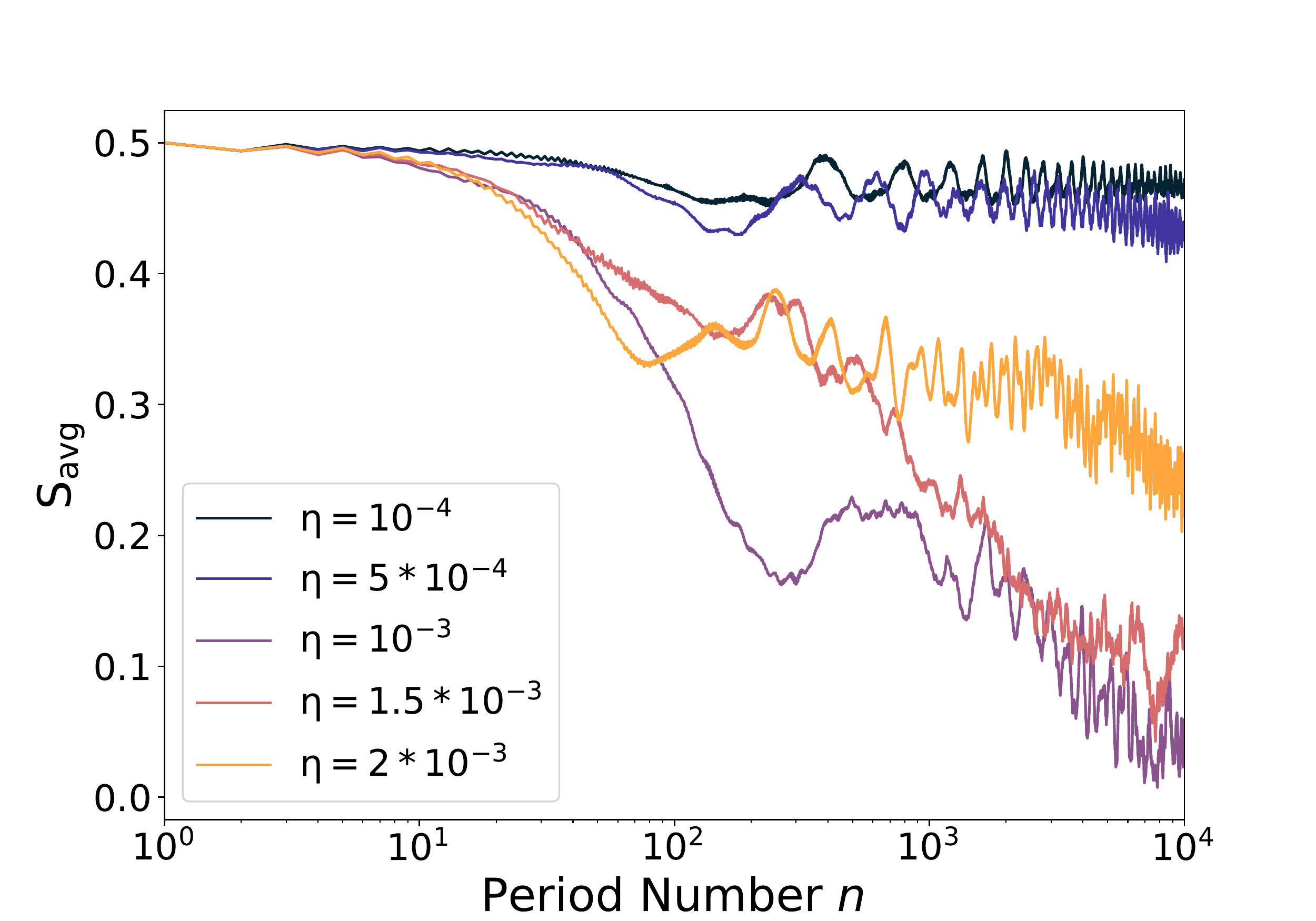}
    \caption{Effect of finite pulse durations on the time-crystalline bahvior. The satellite spin magnetization is shown for a periodically driven central-spin model with isotropic Heisenberg interactions with $J_{x,y}/2\pi=J_z/2\pi=J/2\pi=1$ MHz. We set $B_\mathrm{c}/2\pi$=300 MHz, $B_\mathrm{{sat}}/2\pi$=3 MHz with uniform disorder $\delta B_\mathrm{sat}/2\pi$=0.05 MHz. The Floquet driving error is $e_\mathrm{c,sat}=0.05$, and the driving period is $T=1~\mu$s.}
    \label{fig:pulsedifferenttimesbc300}
\end{figure}

In this section, we investigate the robustness of time-crystalline behavior when we replace the idealized, instantaneous pulses with pulses of finite amplitude and duration. Here, we focus on time-crystalline order that is induced by a large central-spin Zeeman energy with time-dependent driving. A high magnetic field affects the application of single-qubit gates in terms of time and fidelity. 

We consider the case in which finite $\pi$-pulses are implemented via separate AC drives on the central and satellite spins. These drives are chosen to have frequency $B_\mathrm{c}$ for the central spin and $B_\mathrm{{sat}}$ for the satellite spins in order to be on resonance. The central-satellite spin interactions, Eq.~\eqref{eq:hamiltonian}, are present during the application of these pulses, as is of course consistent with experimental implementations. 
We incorporate this finite driving by adding the following terms to the Hamiltonian:
\begin{equation}\label{eq:sxnon}
\begin{aligned}
    &V_0(t)=\frac{\pi(1-e_c)}{\eta T}\cos(B_\mathrm{c}t)S_{x,0},\\
    &V_i(t)=\frac{\pi(1-e_{sat})}{\eta T}\cos(B_\mathrm{{sat}}t)\sum_{i=1}^{N-1}S_{x,i},\\
    & \mathrm{for}\,\,sT\,-\,\eta T \,<\, t\, <\, sT,\, s \in \mathbb{Z}^+. 
\end{aligned}
\end{equation}
With these driving terms included, we study how the subharmonic response is affected by the finite pulse time. The results are shown in Fig.~\ref{fig:pulsedifferentfield}. In this figure, we sweep the central-spin Zeeman energy $B_\mathrm{c}/2\pi$ with the satellite Zeeman energy set to $B_\mathrm{{sat}}=0.01B_\mathrm{c}$, starting from  50 MHz until 300 MHz. Interestingly, we see that the subharmonic response is strongest for different values of $B_\mathrm{c}$ betweeen the central and satellite spins. While the satellite spins are most stable for $B_\mathrm{c}/2\pi=300$ MHz, the central spin exhibits the strongest harmonic response at $B_\mathrm{sat}/2\pi=100$ MHz among the values considered. This is due to the non-monotonic behavior as a function of $B_\mathrm{sat}$ observed in Fig.~\ref{fig:effect_of_Bsat}. 

We also examine the effect of different pulse durations starting from 0.1 ns up to  2 ns, as shown in Fig.~\ref{fig:pulsedifferenttimesbc300}. We fix the Zeeman energies at $B_\mathrm{{c}}/2\pi=300$ MHz and $B_\mathrm{{sat}}/2\pi=0.01B_\mathrm{c}/2\pi=3$ MHz in the presence of rotation error $e_\mathrm{c,sat}=0.05$.  Due to the high central spin magnetic field, the dynamics of the central spin are complicated. That is why we keep our analysis focused on the satellite spins. The figure shows that if the pulses are fast enough, the subharmonic response is achieved even with imperfect time-dependent driving, as in the case of delta-function driving. However, the response quickly decays as the pulse time is increased. It is important to note, though, that the nanosecond timescales considered here are specific to the arbitrary choice of $T=1~\mu$s as the period. The pulse times for which a subharmonic response is visible can be increased by increasing $T$.

\section{Dependence on number of spins}\label{sec:numberspins}

\begin{figure}
\includegraphics[width=\columnwidth]{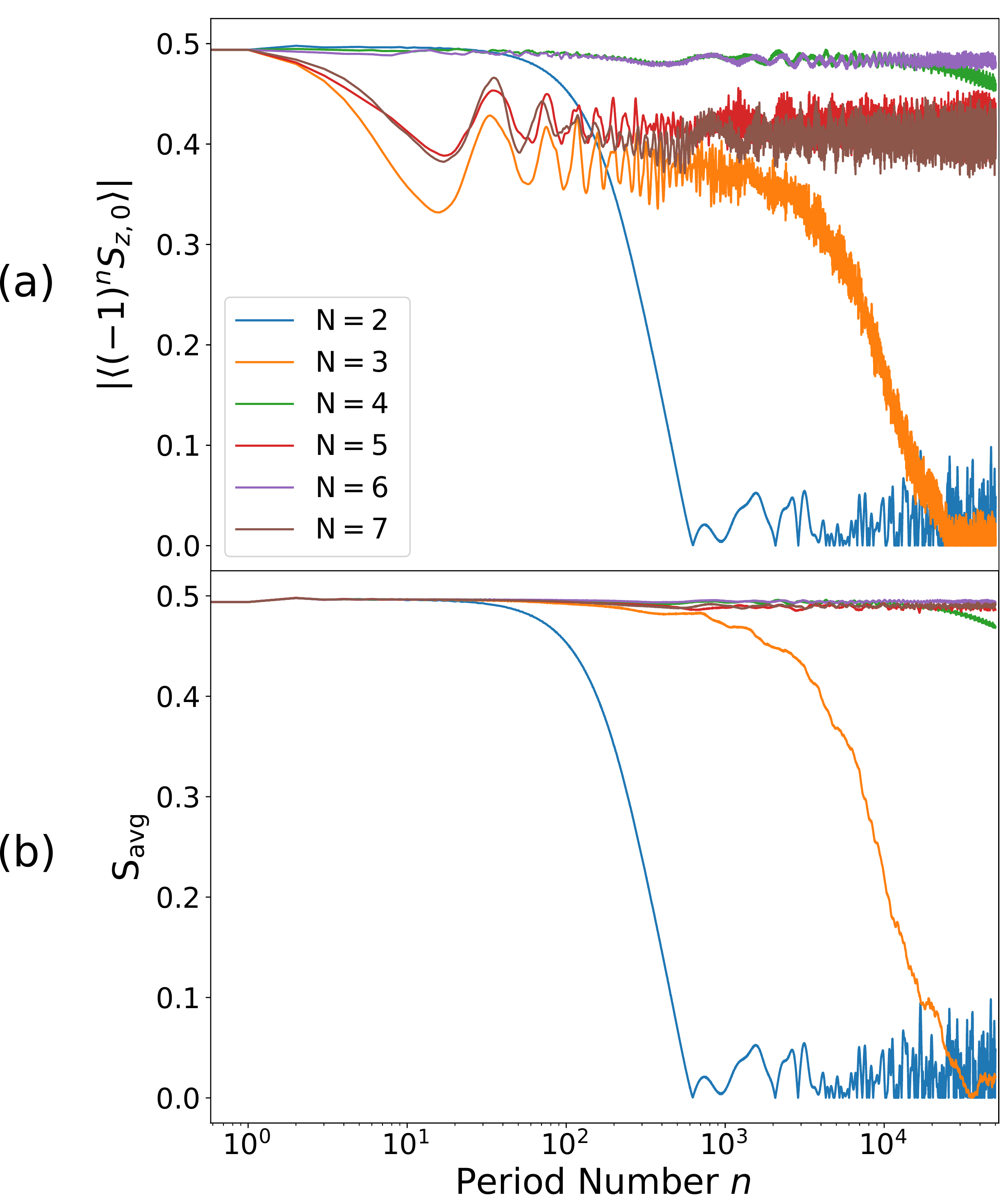}
\caption{(a) Central-spin and (b) satellite-spin magnetizations for a periodically driven central-spin model with isotropic interactions. Results are shown for a varying number $N$ of spins. In each case, the number of satellite spins is $N-1$. The interaction strength is $J_z/2\pi=J_{xy}/2\pi=$1 MHz, the disorder strength is $\delta J/2\pi=0.2$ MHz, the Zeeman energies are $B_\mathrm{c}/2\pi=300$ MHz and $B_\mathrm{sat}=0$, the Floquet driving error is $e_\mathrm{c,sat}=0.05$, and the driving period is $T=1~\mu$s.}\label{fig:changing_N}
\end{figure}

In the main text, we report results for the dynamics of spin expectation values focusing mostly on the case of $N=6$ spins (where there are $N-1$ satellite spins). In this appendix, we study how the time-crystalline behavior depends on the number of spins. We focus on isotropic Heisenberg interactions; we observe similar results when we have anisotropic interactions. In the presence of strong enough disorder, we can see that the mean magnetization of satellite spins stays close to 0.5 for longer times as we increase the number of satellite spins (Fig.~\ref{fig:changing_N}).

It is also evident from the figure that the central-spin magnetization exhibits different behavior depending on whether the total number of spins $N$ is even or odd. We can shed light on this using an effective Hamiltonian that is valid in the limit of large central-spin Zeeman energy, as we now explain.

In the limit where $B_\mathrm{c}$ is very large, we can neglect the flip-flop terms in the Hamiltonian so that it becomes effectively Ising-like:
\begin{equation}\label{eq:hamiltonianIsing}
 \begin{aligned}
   &H_\mathrm{eff}=J_{z,i}\,\,S_{z,0}\sum_{i=1}^{N-1}S_{z,i}+B_\mathrm{c}S_{z,0}.\\
 \end{aligned}
\end{equation}
In what follows, we use this effective Hamiltonian to simulate the dynamics after two periods to see why there is a decrease in the central-spin magnetization for an odd number of spins compared to an even number. First, we focus on the presence of central-spin $\pi$ pulse driving error $e_\mathrm{c}$, and we set $e_\mathrm{sat}=0$. The system evolves under the Floquet operator $U=\prod_{n=1}^\mathcal{N} (U_\pi U_{H_\mathrm{eff}})$ where $U_\pi = \prod_i e^{-i\pi(1-0) S_{x,i}} e^{-i\pi(1-e_{\mathrm{c}}) S_{x,0}}$. We simulate our system for an even number of periods ( $\mathcal{N}$=2,4,6....). We calculate the time-evolved state after $\mathcal{N}$ number of periods for $N=$3 and 4 spins: 
\begin{equation}
    \ket{\Psi(t)_{\rm{N}=3}}\propto  \begin{pmatrix}
    \alpha_{1}\cos(\frac{\mathcal{N}}{2}e_\mathrm{c})+i\alpha_{5}\sin(\frac{\mathcal{N}}{2}e_\mathrm{c})\\
     \alpha_{2}\cos(\frac{\mathcal{N}}{2}e_\mathrm{c})+i\alpha_{6}\sin(\frac{\mathcal{N}}{2}e_\mathrm{c})\\
    \alpha_{3}\cos(\frac{\mathcal{N}}{2}e_\mathrm{c})+i\alpha_{7}\sin(\frac{\mathcal{N}}{2}e_\mathrm{c})\\
    \alpha_{4}\cos(\frac{\mathcal{N}}{2}e_\mathrm{c})+i\alpha_{8}\sin(\frac{\mathcal{N}}{2}e_\mathrm{c})\\
    \alpha_{5}\cos(\frac{\mathcal{N}}{2}e_\mathrm{c})+i\alpha_{1}\sin(\frac{\mathcal{N}}{2}e_\mathrm{c})\\
    \alpha_{6}\cos(\frac{\mathcal{N}}{2}e_\mathrm{c})+i\alpha_{2}\sin(\frac{\mathcal{N}}{2}e_\mathrm{c})\\
    \alpha_{7}\cos(\frac{\mathcal{N}}{2}e_\mathrm{c})+i\alpha_{3}\sin(\frac{\mathcal{N}}{2}e_\mathrm{c})\\
    \alpha_{8}\cos(\frac{\mathcal{N}}{2}e_\mathrm{c})+i\alpha_{4}\sin(\frac{\mathcal{N}}{2}e_\mathrm{c})\\
    
     \end{pmatrix}
\end{equation}
starting from an initial state $\ket{\Psi(0)_{\rm{N}=3}}=( \alpha_{1}, \alpha_{2}, \alpha_{3}, \alpha_{4}, \alpha_{5}, \alpha_{6}, \alpha_{7}, \alpha_{8})$.
On the other hand, in the case of an even number of spins ($N=4$), we can see the period doubling effect in the presence of a $\pi$ pulse driving error on the central spin. If we start with an initial state 
$\ket{\Psi(0)_{\rm{N}=4}}=( \beta_{1}, \beta_{2}, \beta_{3}, \beta_{4}, \beta_{5}, \beta_{6}, \beta_{7}, \beta_{8},\beta_{9}, \beta_{10}, \beta_{11}, \beta_{12}, \beta_{13}, \beta_{14}, \beta_{15},$ $ \beta_{16})$, we perfectly recover the initial state after an even number $\mathcal{N}$ of periods: $\ket{\Psi(t)_{\rm{N}=4}}\propto\ket{\Psi(0)_{\rm{N}=4}}$. The emergence of perfect period doubling is due to a specific many-body interaction strength ($J_{z,i}/2\pi=J/2\pi$=1 MHz) and a specific value of the on-site central-spin Zeeman splitting ($B_\mathrm{c}/2\pi$=300 MHz).

If we vary the Zeeman energy while keeping constant the many-body interaction strength ($J/2\pi=1$ MHz) and the interaction time ($T=1~\mu$s), we will observe the following state after $\mathcal{N}=2$ periods. (We focus on the first period doubling period ($\mathcal{N}$=2) because it is easier to identify why we have a perfect period doubling effect in the presence of central-spin $\pi$ pulse driving error.) Starting with the same initial states for $N=3, 4$ spins, respectively, the resulting states are 
\begin{eqnarray}
&&\ket{\Psi(t)_{N=3,\mathcal{N}=2,J/2\pi=1\,\, \rm{MHz}}}=\nonumber\\&&
    \!\!\!\!\!\!\!\!\!\!\colvec[.85]{    
    \frac{\alpha_{1}}{2}((-1+e^{-i B_\mathrm{c}})-(1+e^{-i  B_\mathrm{c}})\cos(e_\mathrm{c}))-i\frac{\alpha_{5}}{2}(1+e^{i B_\mathrm{c}})\sin(e_\mathrm{c}))\\
     \frac{\alpha_{2}}{2}((-1+e^{-i B_\mathrm{c}})-(1+e^{-i  B_\mathrm{c}})\cos(e_\mathrm{c}))-i\frac{\alpha_{6}}{2}(1+e^{i B_\mathrm{c}})\sin(e_\mathrm{c}))\\
     \frac{\alpha_{3}}{2}((-1+e^{-i B_\mathrm{c}})-(1+e^{-i  B_\mathrm{c}})\cos(e_\mathrm{c}))-i\frac{\alpha_{7}}{2}(1+e^{i B_\mathrm{c}})\sin(e_\mathrm{c}))\\
     \frac{\alpha_{4}}{2}((-1+e^{-i B_\mathrm{c}})-(1+e^{-i  B_\mathrm{c}})\cos(e_\mathrm{c}))-i\frac{\alpha_{8}}{2}(1+e^{i B_\mathrm{c}})\sin(e_\mathrm{c}))\\
    \frac{\alpha_{5}}{2}((-1+e^{i B_\mathrm{c}})-(1+e^{i  B_\mathrm{c}})\cos(e_\mathrm{c}))-i\frac{\alpha_{1}}{2}(1+e^{-i B_\mathrm{c}})\sin(e_\mathrm{c}))\\
    \frac{\alpha_{6}}{2}((-1+e^{i B_\mathrm{c}})-(1+e^{i  B_\mathrm{c}})\cos(e_\mathrm{c}))-i\frac{\alpha_{2}}{2}(1+e^{-i B_\mathrm{c}})\sin(e_\mathrm{c}))\\
    \frac{\alpha_{7}}{2}((-1+e^{i B_\mathrm{c}})-(1+e^{i  B_\mathrm{c}})\cos(e_\mathrm{c}))-i\frac{\alpha_{3}}{2}(1+e^{-i B_\mathrm{c}})\sin(e_\mathrm{c}))\\
   \frac{\alpha_{8}}{2}((-1+e^{i B_\mathrm{c}})-(1+e^{i  B_\mathrm{c}})\cos(e_\mathrm{c}))-i\frac{\alpha_{4}}{2}(1+e^{-i B_\mathrm{c}})\sin(e_\mathrm{c}))\\
    },\nonumber\\&&
\end{eqnarray}
and
\begin{eqnarray}
&&\ket{\Psi(t)_{N=4,\mathcal{N}=2,J/2\pi=1\,\, \rm{MHz}}}=\nonumber\\&&
    \!\!\!\!\!\!\!\!\!\!\colvec[.9]{
    \frac{\beta_{1}}{2}((1+e^{-i B_\mathrm{c}})-(-1+e^{-i  B_\mathrm{c}})\cos e_\mathrm{c})-i\frac{\beta_{9}}{2}(-1+e^{i B_\mathrm{c}})\sin e_\mathrm{c})\\
    \frac{\beta_{2}}{2}((1+e^{-i B_\mathrm{c}})-(-1+e^{-i  B_\mathrm{c}})\cos e_\mathrm{c})-i\frac{\beta_{10}}{2}(-1+e^{i B_\mathrm{c}})\sin e_\mathrm{c})\\
    \frac{\beta_{3}}{2}((1+e^{-i B_\mathrm{c}})-(-1+e^{-i  B_\mathrm{c}})\cos e_\mathrm{c})-i\frac{\beta_{11}}{2}(-1+e^{i B_\mathrm{c}})\sin e_\mathrm{c})\\
    \frac{\beta_{4}}{2}((1+e^{-i B_\mathrm{c}})-(-1+e^{-i  B_\mathrm{c}})\cos e_\mathrm{c})-i\frac{\beta_{12}}{2}(-1+e^{i B_\mathrm{c}})\sin e_\mathrm{c})\\
    \frac{\beta_{5}}{2}((1+e^{-i B_\mathrm{c}})-(-1+e^{-i  B_\mathrm{c}})\cos e_\mathrm{c})-i\frac{\beta_{13}}{2}(-1+e^{i B_\mathrm{c}})\sin e_\mathrm{c})\\
    \frac{\beta_{6}}{2}((1+e^{-i B_\mathrm{c}})-(-1+e^{-i  B_\mathrm{c}})\cos e_\mathrm{c})-i\frac{\beta_{14}}{2}(-1+e^{i B_\mathrm{c}})\sin e_\mathrm{c})\\
    \frac{\beta_{7}}{2}((1+e^{-i B_\mathrm{c}})-(-1+e^{-i  B_\mathrm{c}})\cos e_\mathrm{c})-i\frac{\beta_{15}}{2}(-1+e^{i B_\mathrm{c}})\sin e_\mathrm{c})\\
    \frac{\beta_{8}}{2}((1+e^{-i B_\mathrm{c}})-(-1+e^{-i  B_\mathrm{c}})\cos e_\mathrm{c})-i\frac{\beta_{16}}{2}(-1+e^{i B_\mathrm{c}})\sin e_\mathrm{c})\\
    \frac{\beta_{9}}{2}((1+e^{i B_\mathrm{c}})-(-1+e^{i  B_\mathrm{c}})\cos e_\mathrm{c})-i\frac{\beta_{1}}{2}(-1+e^{-i B_\mathrm{c}})\sin e_\mathrm{c})\\
    \frac{\beta_{10}}{2}((1+e^{i B_\mathrm{c}})-(-1+e^{i  B_\mathrm{c}})\cos e_\mathrm{c})-i\frac{\beta_{2}}{2}(-1+e^{-i B_\mathrm{c}})\sin e_\mathrm{c})\\
    \frac{\beta_{11}}{2}((1+e^{i B_\mathrm{c}})-(-1+e^{i  B_\mathrm{c}})\cos e_\mathrm{c})-i\frac{\beta_{3}}{2}(-1+e^{-i B_\mathrm{c}})\sin e_\mathrm{c})\\
    \frac{\beta_{12}}{2}((1+e^{i B_\mathrm{c}})-(-1+e^{i  B_\mathrm{c}})\cos e_\mathrm{c})-i\frac{\beta_{4}}{2}(-1+e^{-i B_\mathrm{c}})\sin e_\mathrm{c})\\
    \frac{\beta_{13}}{2}((1+e^{i B_\mathrm{c}})-(-1+e^{i  B_\mathrm{c}})\cos e_\mathrm{c})-i\frac{\beta_{5}}{2}(-1+e^{-i B_\mathrm{c}})\sin e_\mathrm{c})\\
    \frac{\beta_{14}}{2}((1+e^{i B_\mathrm{c}})-(-1+e^{i  B_\mathrm{c}})\cos e_\mathrm{c})-i\frac{\beta_{6}}{2}(-1+e^{-i B_\mathrm{c}})\sin e_\mathrm{c})\\
    \frac{\beta_{15}}{2}((1+e^{i B_\mathrm{c}})-(-1+e^{i  B_\mathrm{c}})\cos e_\mathrm{c})-i\frac{\beta_{7}}{2}(-1+e^{-i B_\mathrm{c}})\sin e_\mathrm{c})\\
    \frac{\beta_{16}}{2}((1+e^{i B_\mathrm{c}})-(-1+e^{i  B_\mathrm{c}})\cos e_\mathrm{c})-i\frac{\beta_{8}}{2}(-1+e^{-i B_\mathrm{c}})\sin e_\mathrm{c})\\
    }.\nonumber\\&&
\end{eqnarray}
If we instead vary the interaction strength $J$ and while keeping constant the central-spin Zeeman energy ($B_\mathrm{c}/2\pi=300$ MHz), we obtain the following results:
\begin{eqnarray}
&&\ket{\Psi(t)_{N=3,\mathcal{N}=2,B_\mathrm{c}/2\pi=300\,\, \rm{MHz}}}=\nonumber\\&&
    \!\!\!\!\!\!\!\!\!\!\colvec[.9]{
    \frac{\alpha_{1}}{2}(1-e^{-i J}-\cos(e_\mathrm{c})-e^{-i J}\cos(e_\mathrm{c}))-i\frac{\alpha_{5}}{2}(1+e^{i J})\sin(e_\mathrm{c}))\\
     -\alpha_{2}\cos(e_\mathrm{c})-i\alpha_{6}\sin(e_\mathrm{c})\\
    -\alpha_{3}\cos(e_\mathrm{c})-i\alpha_{7}\sin(e_\mathrm{c})\\
     \frac{\alpha_{4}}{2}(1-e^{i J}-\cos(e_\mathrm{c})-e^{i J}\cos(e_\mathrm{c}))-i\frac{\alpha_{8}}{2}(1+e^{-i J})\sin(e_\mathrm{c}))\\
    \frac{\alpha_{5}}{2}(1-e^{i J}-\cos(e_\mathrm{c})-e^{i J}\cos(e_\mathrm{c}))-i\frac{\alpha_{1}}{2}(1+e^{-i J})\sin(e_\mathrm{c}))\\
    -\alpha_{6}\cos(e_\mathrm{c})-i\alpha_{2}\sin(e_\mathrm{c})\\
    -\alpha_{7}\cos(e_\mathrm{c})-i\alpha_{3}\sin(e_\mathrm{c})\\
    \frac{\alpha_{8}}{2}(1-e^{-i J}-\cos(e_\mathrm{c})-e^{-i J}\cos(e_\mathrm{c}))-i\frac{\alpha_{4}}{2}(1+e^{i J})\sin(e_\mathrm{c}))\\
     },\nonumber\\&&
\end{eqnarray}
and
\begin{eqnarray}\label{eq:n4j}
&&\ket{\Psi(t)_{N=4,\mathcal{N}=2, B_\mathrm{c}/2\pi=300\,\, \rm{MHz}}}=\nonumber\\&&
    \!\!\!\!\!\!\!\!\!\!\colvec[.9]{
    \frac{\beta_{1}}{2}(-1+e^{-\frac{3iJ}{2}}+\cos e_\mathrm{c}+e^{-\frac{3iJ}{2}}\cos e_\mathrm{c})+i\frac{\beta_{9}}{2}(1+e^{\frac{3iJ}{2}})\sin e_\mathrm{c})\\
    \frac{\beta_{2}}{2}(-1+e^{-\frac{iJ}{2}}+\cos e_\mathrm{c}+e^{-\frac{iJ}{2}}\cos e_\mathrm{c})+i\frac{\beta_{10}}{2}(1+e^{\frac{iJ}{2}})\sin e_\mathrm{c})\\
    \frac{\beta_{3}}{2}(-1+e^{-\frac{iJ}{2}}+\cos e_\mathrm{c}+e^{-\frac{iJ}{2}}\cos e_\mathrm{c})+i\frac{\beta_{11}}{2}(1+e^{\frac{iJ}{2}})\sin e_\mathrm{c})\\
    \frac{\beta_{4}}{2}(-1+e^{+\frac{iJ}{2}}+\cos e_\mathrm{c}+e^{+\frac{iJ}{2}}\cos e_\mathrm{c})+i\frac{\beta_{12}}{2}(1+e^{-\frac{iJ}{2}})\sin e_\mathrm{c})\\
    \frac{\beta_{5}}{2}(-1+e^{-\frac{iJ}{2}}+\cos e_\mathrm{c}+e^{-\frac{iJ}{2}}\cos e_\mathrm{c})+i\frac{\beta_{13}}{2}(1+e^{\frac{iJ}{2}})\sin e_\mathrm{c})\\
    \frac{\beta_{6}}{2}(-1+e^{\frac{iJ}{2}}+\cos e_\mathrm{c}+e^{\frac{iJ}{2}}\cos e_\mathrm{c})+i\frac{\beta_{14}}{2}(1+e^{-\frac{iJ}{2}})\sin e_\mathrm{c})\\
    \frac{\beta_{7}}{2}(-1+e^{\frac{iJ}{2}}+\cos e_\mathrm{c}+e^{\frac{iJ}{2}}\cos e_\mathrm{c})+i\frac{\beta_{15}}{2}(1+e^{-\frac{iJ}{2}})\sin e_\mathrm{c})\\
    \frac{\beta_{8}}{2}(-1+e^{\frac{3iJ}{2}}+\cos e_\mathrm{c}+e^{\frac{3iJ}{2}}\cos e_\mathrm{c})+i\frac{\beta_{16}}{2}(1+e^{-\frac{3iJ}{2}})\sin e_\mathrm{c})\\
    \frac{\beta_{9}}{2}(-1+e^{\frac{3iJ}{2}}+\cos e_\mathrm{c}+e^{\frac{3iJ}{2}}\cos e_\mathrm{c})+i\frac{\beta_{1}}{2}(1+e^{-\frac{3iJ}{2}})\sin e_\mathrm{c})\\
    \frac{\beta_{10}}{2}(-1+e^{\frac{iJ}{2}}+\cos e_\mathrm{c}+e^{\frac{iJ}{2}}\cos e_\mathrm{c})+i\frac{\beta_{2}}{2}(1+e^{-\frac{iJ}{2}})\sin e_\mathrm{c})\\
    \frac{\beta_{11}}{2}(-1+e^{\frac{iJ}{2}}+\cos e_\mathrm{c}+e^{\frac{iJ}{2}}\cos e_\mathrm{c})+i\frac{\beta_{3}}{2}(1+e^{-\frac{iJ}{2}})\sin e_\mathrm{c})\\
    \frac{\beta_{12}}{2}(-1+e^{-\frac{iJ}{2}}+\cos e_\mathrm{c}+e^{-\frac{iJ}{2}}\cos e_\mathrm{c})+i\frac{\beta_{4}}{2}(1+e^{\frac{iJ}{2}})\sin e_\mathrm{c})\\
    \frac{\beta_{13}}{2}(-1+e^{\frac{iJ}{2}}+\cos e_\mathrm{c}+e^{\frac{iJ}{2}}\cos e_\mathrm{c})+i\frac{\beta_{5}}{2}(1+e^{-\frac{iJ}{2}})\sin e_\mathrm{c})\\
    \frac{\beta_{14}}{2}(-1+e^{-\frac{iJ}{2}}+\cos e_\mathrm{c}+e^{-\frac{iJ}{2}}\cos e_\mathrm{c})+i\frac{\beta_{6}}{2}(1+e^{\frac{iJ}{2}})\sin e_\mathrm{c})\\
    \frac{\beta_{15}}{2}(-1+e^{-\frac{iJ}{2}}+\cos e_\mathrm{c}+e^{-\frac{iJ}{2}}\cos e_\mathrm{c})+i\frac{\beta_{7}}{2}(1+e^{\frac{iJ}{2}})\sin e_\mathrm{c})\\
    \frac{\beta_{16}}{2}(-1+e^{-\frac{3iJ}{2}}+\cos e_\mathrm{c}+e^{-\frac{3iJ}{2}}\cos e_\mathrm{c})+i\frac{\beta_{8}}{2}(1+e^{\frac{3iJ}{2}})\sin e_\mathrm{c})\\
    }.\nonumber\\&&
\end{eqnarray}

What is needed to make the return probability equal to one after an even number of periods is different in each case. As we can see from the coefficients of the $N=3$ case, if we assume that $(1+e^{\pm i B_\mathrm{c}})=0$, we can observe a perfect period doubling effect. On the other hand, in the case of an even number of spins ($N=4$), if we eliminate the term $(1-e^{\pm i B_\mathrm{c}})$, we achieve unit probability every two periods:
\begin{equation}
    \begin{aligned}
        &N=3:\,\,(1+e^{\pm i B_\mathrm{c}})=0\xrightarrow{} e^{\pm i B_\mathrm{c}}=e^{i(2x\pi +\pi)}\xrightarrow{} \\
        &|B_\mathrm{c}/2\pi|= (x+1/2)\,\, \rm{MHz}\\
        &N=4:\,\,(1-e^{\pm i B_\mathrm{c}})=0\xrightarrow{} e^{\pm i B_\mathrm{c}}=e^{i2x\pi}\xrightarrow{}\\ &|B_\mathrm{c}/2\pi|= x \,\,\rm{MHz}\\
    \end{aligned}
\end{equation}
where $x\in \mathbb{Z}$. This is why we observe a stronger subharmonic response for an even number of spins. In our simulations, we assumed a central-spin Zeeman energy of $B_\mathrm{c}/2\pi=300$ MHz. However, if we change this to $B_\mathrm{c}/2\pi=300.5$ MHz, we effectively swap the even-odd behavior. In this case, an odd number of spins will exhibit a stronger subharmonic response compared to an even number of spins.

In addition to the central-spin Zeeman energy, the interaction strength $J$ also plays an important role. In this case, we fix the central-spin Zeeman energy to $B_\mathrm{c}/2\pi=300$ MHz. To eliminate unwanted coefficients in the final state, we have to choose specific values of $J$ to perfectly retrieve the state after every two periods. In particular, we have to eliminate the factors involving $\cos e_\mathrm{c}\,\, \rm{and}\,\, \sin e_\mathrm{c}$ in the many-body time-evolved state:
\begin{equation}
    \begin{aligned}
        &N=3:\,\,(1+e^{\pm i J})=0\xrightarrow{} e^{i\pm J}=e^{i(2x\pi +\pi)}\xrightarrow{} \\
        &|J|/2\pi= (x+1/2)\,\, \rm{MHz}\\
        &\rm{N=4:}\,\,(1+e^{i\pm \frac{J}{2}})=0\xrightarrow{} e^{i\pm \frac{J}{2}}=e^{i(2x\pi +\pi)}\\
        &\xrightarrow{}|J/2\pi|= (2x+1) \,\,\rm{MHz}\\
        &\rm{And}\\
        &N=4:\,\,(1+e^{i\pm \frac{3J}{2}})=0\xrightarrow{} e^{i\pm \frac{3J}{2}}=e^{i(2x\pi +\pi)}\\
        &\xrightarrow{}|J/2\pi|= \frac{2x+1}{3} \,\,\rm{MHz}\\
    \end{aligned}
\end{equation}

\section{Insensitivity to the initial state}\label{sec:initial}

Here, we examine how the decay of the return probability depends on the initial state. Fixing the number of spins to $N=6$, we compute the number of Floquet cycles over which the return probability remains above 0.95 when each of the $2^{6}$ basis states is taken as the initial state. The results for isotropic interactions are shown in Fig.~\ref{fig:isoj1initial} as a function of the $\pi$ pulse error. We calculate the final Floquet cycle in the presence of interaction disorder $\delta J/2\pi=0.2$ MHz and with central- and satellite-spin Zeeman splittings of ${B_\mathrm{c}}/2\pi$ = 300~MHz and ${B_\mathrm{sat}}=0$, respectively. We see that the time crystal phase region exhibits a weak dependence on the initial state. The behavior is not significantly different for the case of anisotropic interactions where $J_z\gg J_{xy}$.
\begin{figure}
    \includegraphics[width=\columnwidth]{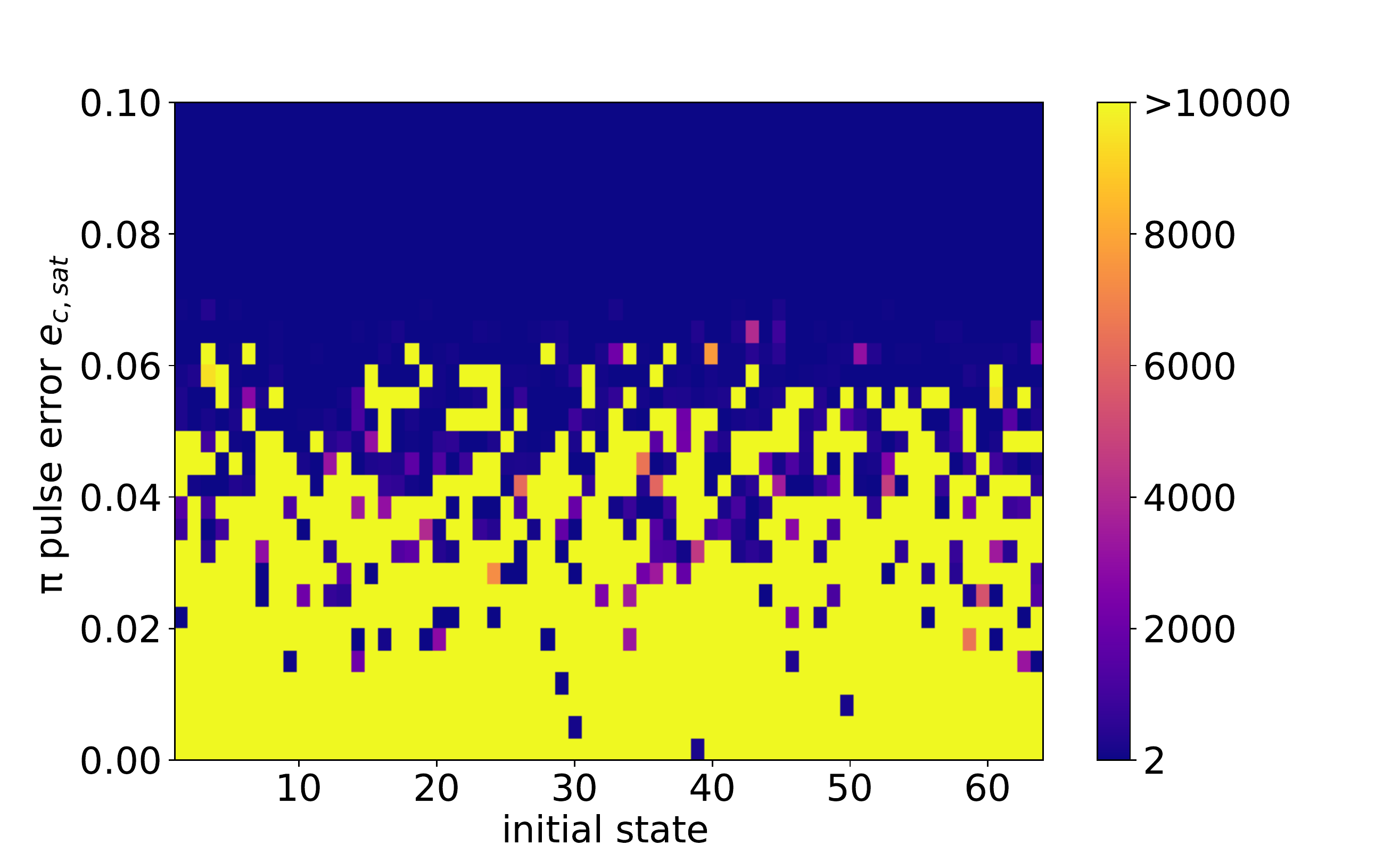}    
    \caption{The number of
    Floquet cycles (color bar) over which the return probability of
    the full central-spin system evolves stroboscopically (see main
    text for precise definition) as a function of the pulse error $\rm{e_{c,sat}}$ and for 64 different initial states, corresponding to the $2^6$ distinct basis states. We assume isotropic interactions of strength $J_z/2\pi=J_{xy}/2\pi=1$ MHz. Here, $B_\mathrm{c}/2\pi=300$~MHz, $B_\mathrm{sat}=0$, $\delta J/2\pi=0.2$ MHz, $e_\mathrm{c,sat}=0.05$, $T=1~\mu$s.}
    \label{fig:isoj1initial}
\end{figure}

\section{Mixed states in satellite spins}\label{sec:mixed}
Here, we consider the impact on the subharmonic response when the initial state is not pure. Specifically, we will assume that the satellite spins undergo a depolarization channel. 
\begin{equation}
    \mathcal{E}(\rho) = (1- p )\rho +\frac{p}{2} I \ ,
    \label{eq:dep_error}
\end{equation}
Even in the case of a mixed state, our initial state is stabilized. As we show in the following figures, even in the presence of a high depolarization rate, the state is stabilized in presence of imperfect $\pi$-pulse driving. We give a simple example assuming that the initial state is prepared by applying the above depolarization channel on each satellite spin of the state $\ket{\Psi(0)}=\ket{\uparrow\uparrow\downarrow\uparrow\downarrow\uparrow}$. Due to the presence of the depolarization error in the satellite spins, the updated initial state becomes:
\begin{eqnarray}
    &&\rho_{initial}=\nonumber\\&&\ket{\uparrow}\bra{\uparrow}\mathcal{E}(\ket{\uparrow}\bra{\uparrow})\mathcal{E}(\ket{\downarrow}\bra{\downarrow}) \mathcal{E}(\ket{\uparrow}\bra{\uparrow})\mathcal{E}(\ket{\downarrow}\bra{\downarrow}) \mathcal{E}(\ket{\uparrow}\bra{\uparrow}),\nonumber\\
\end{eqnarray}
where the same depolarization rate $p$ is used for all satellite spins. In the following figures, we examine how the stabilization of the mixed state is preserved. We investigate both methods for inducing time-crystalline behavior: using a large Zeeman splitting mismatch, and using 30 H2I pulses. We show that the satellite's average magnetization is stabilized. However, in the case of the central spin, even if it starts in a pure state, its interaction with satellitespins prepared in a mixed state leads to a destabilization of the subharmonic response.

\begin{figure}[tbph]
    \centering
    \includegraphics[scale=0.28]{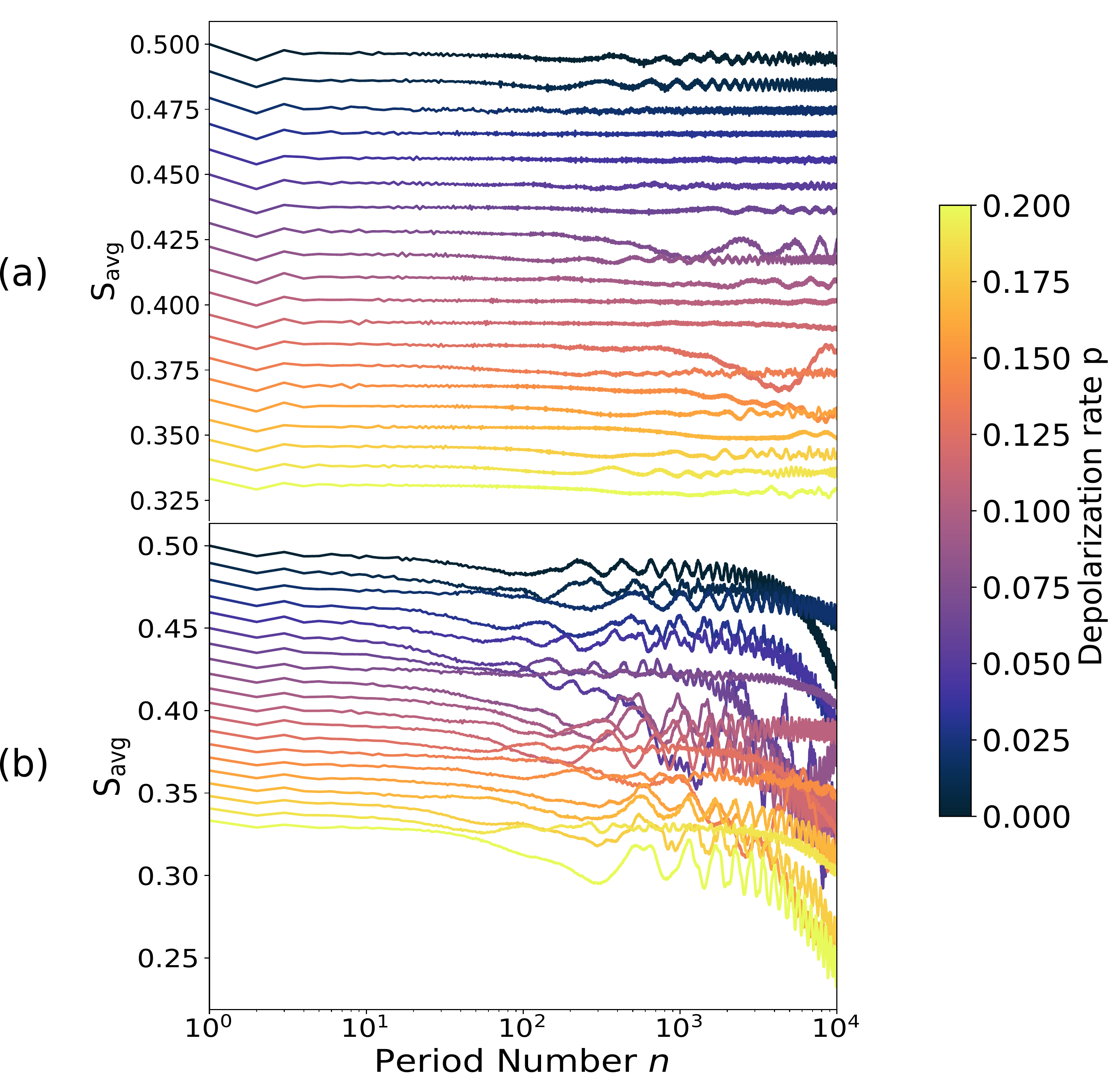}
    \caption{Effect of starting from a mixed satellite spin state obtained by subjecting each satellite spin to a depolarizing channel characterized by rate $p$. The satellite spin magnetization in a periodically driven central-spin model with isotropic Heisenberg interactions with $J_{x,y}/2\pi=J_z/2\pi=J/2\pi=1$ MHz is shown. The time-crystalline behavior is induced with (a) a large Zeeman splitting mismatch with $B_\mathrm{c}/2\pi$=300 MHz and $B_\mathrm{{sat}}$=0 MHz, and (b) 30 additional perfect H2I pulses applied to the central spin every Floquet period. In both panels, the $\pi$-pulse driving error is $e_\mathrm{c,sat}=0.05$.}
    \label{fig:savgmixeddepolarization}
\end{figure}
Figure~\ref{fig:savgmixeddepolarization} shows that a clear submharmonic response remains evident in the satellite spins even when they are initialized in a mixed state. As the depolarization rate increases, the satellite spin magnetization has a diminished amplitude, but it continues to exhibit period doubling. In Fig.~\ref{fig:savgmixeddepolarization}(b), we also see that the subharmonic response becomes unstable on long timescales due to the relatively small number of H2I pulses used in this case (m=30). Here, we do not include the central-spin magnetization, because it does not exhibit any subharmonic response in the presence of a nonzero depolarization of the nuclear spins. 

\section{Resonances of central-spin model}\label{sec:resonances}

As we saw from the phase diagrams, when we swept the interaction strength, we observed regions with no subharmonic response. For example, whereas $J/2\pi=1$ MHz produces the strongest subharmonic response in our system, in the case of $J/2\pi=2$ MHz, the time crystal-like phase is destroyed. We know from theoretical investigations~\cite{khemani2019briefhistory} that to achieve a subharmonic response, the many-body Floquet spectrum must exhibit particular properties. Specifically, for period doubling it has been shown that the eigenvalues of the Floquet operator come in antipodal pairs. Here, the Floquet operator is $U_{F}(T)=U_\pi e^{-iHT}$, where $U_\pi = \prod_i e^{-i\pi(1-e_{\mathrm{sat}}) S_{x,i}} e^{-i\pi(1-e_{\mathrm{c}}) S_{x,0}}$.
Using the Hamiltonian in Eq.~\eqref{eq:hamiltonian} with $B_\mathrm{c}/2\pi$=300 MHz, $B_\mathrm{sat}$=0, we compute the eigenvalues of $U_F(T)$ for several values of $J_z=J_{xy}=J$. The results are shown in Fig.~\ref{fig:eigenvaluesevolution}, where it is evident that the eigenvalues for $J/2\pi=1,3$ MHz (black and green circles in the figure) come in antipodal pairs, while those for $J/2\pi=2,4$ MHz (red and blue circles) do not.

In the figure, we show results for $N=6$ spins. The behavior for an odd number of spins is similar to the even number case provided we take $B_c^{\rm{odd}}/2\pi=B_c^{\rm{even}}/2\pi+0.5$ MHz. We again have antipodal pairs of eigenvalues in the case of $J/2\pi=1, 3$ MHz and no antipodal pairs in the case of $J/2\pi=2, 4$ MHz. 
\begin{figure}[tbph]
    \centering
    \includegraphics[width=\columnwidth]{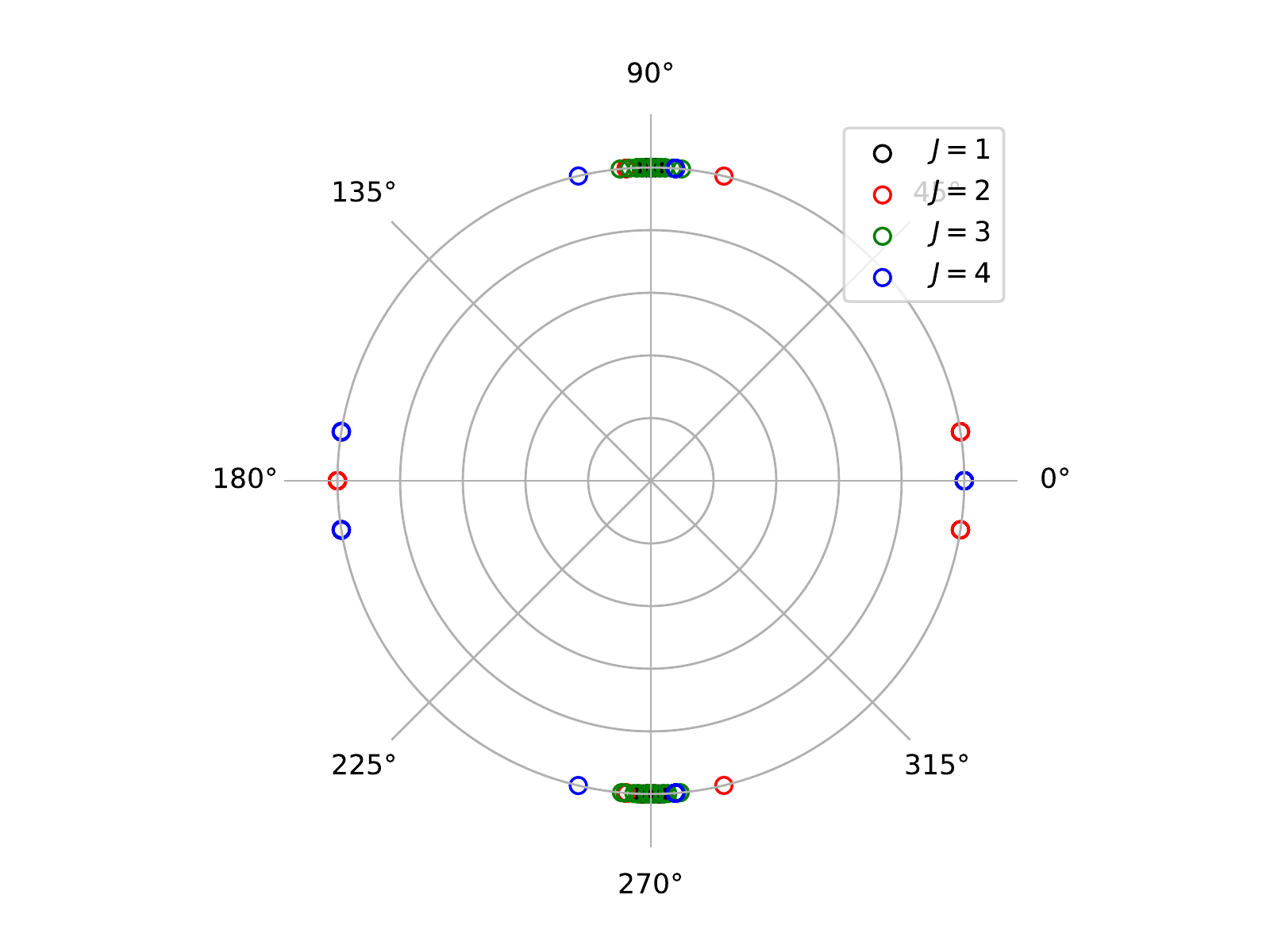}
    \caption{Eigenvalues of the Floquet operator $U_F(T)$ for four different values of the total interaction strength $J_z=J_{xy}=J$ in the case of isotropic interactions in a central-spin system with $N=6$ spins. Here, $B_\mathrm{c}/2\pi$=300 MHz, $B_\mathrm{sat}$=0, $e_\mathrm{c,sat}$=0.05, and $T=1~\mu$s.}
    \label{fig:eigenvaluesevolution}
\end{figure}
To keep the perfect initial state after an even number of periods, we have to apply a specific type of condition to the many-body interaction strength $J$. In the case of even many-body interaction $J/2\pi$ (2, 4 MHz), this means that:
\begin{equation}
 \begin{aligned}
    &(-1+e^{-\frac{iJ}{2}})=0\xrightarrow{}e^{i\pm \frac{J}{2}}=e^{i(2x\pi+2\pi)}\xrightarrow{}\\
    &|J/2\pi|= 2(x+1)\,\, \rm{MHz}\\
\end{aligned}
\end{equation}
As we can see from Eq.~\eqref{eq:n4j}, in the case $J/2\pi=2, 4$ MHz where there is no factor $(-1+e^{-\frac{iJ}{2}})$, there is no absorption of the imperfect driving due to the presence of the factors with $\cos e_\mathrm{c}$ and $\sin e_\mathrm{c}$. However, in the case of $J/2\pi=1, 3$ MHz, the factors of $(1+e^{-\frac{iJ}{2}})$ are eliminated, absorbing the $\pi$ pulse driving with the many-body interaction.

\section{H2I pulse analysis}\label{sec:h2i_pulses}

In this appendix, we provide further analysis of the effect of H2I pulses, and we calculate the dependence of the central spin magnetization on the number of H2I pulses. In the H2I approach, the evolution operator after $m$ H2I pulses can be written as follows:
\begin{equation}
 U_\mathrm{H2I}(T)=[e^{i\pi S_{z,0}(1-e_\mathrm{z})}U_H(T/m)]^{m}.  
\end{equation}
Using the Baker–Campbell–Hausdorff formula we can approximate the application of the two operators by merging them with their commutator. We just keep the first three terms of the series, so we have:
\begin{equation}
    e^{A}e^{B}\approx e^{A+B+\frac{1}{2}[A,B]},
\end{equation}
where
\begin{eqnarray}
A&=&i\pi S_{z,0}(1-e_\mathrm{z}),\nonumber\\
B&=&-iJ\frac{t}{m}\left(\sum_{i=1}^{N-1}S_{x,0}S_{x,i}+S_{y,0}S_{y,i} +S_{z,0}S_{z,i}\right).\nonumber\\
\end{eqnarray}
The commutator $[A,B]$ is readily computed: 
\begin{equation}
\begin{aligned}
    &[A,B]=\left[\alpha S_{z,0}, \beta \left(\sum_{i=1}^{N-1}S_{x,0}S_{x,i}+S_{y,0}S_{y,i} +S_{z,0}S_{z,i}\right)\right]\\
    &=\left[\alpha S_{z,0},\beta S_{x,0}\sum_{i=1}^{N-1}S_{x,i}\right]+\left[\alpha S_{z,0},\beta S_{y,0}\sum_{i=1}^{N-1}S_{y,i}\right]\\
    &=\alpha\beta[S_{z,0}, S_{x,0}]\sum_{i=1}^{N-1}S_{x,i}+\alpha\beta[S_{z,0},S_{y,0}]\sum_{i=1}^{N-1}S_{y,i}\\
    &=\alpha\beta i S_{y,0} \sum_{i=1}^{N-1}S_{x,i}+\alpha\beta (-i) S_{x,0}\sum_{i=1}^{N-1}S_{y,i}\\
    &=\alpha\beta \left(\frac{S_{+,0}-S_{-,0}}{2}\sum_{i=1}^{N-1}S_{x,i}-i\frac{S_{+,0}+S_{-,0}}{2}\sum_{i=1}^{N-1}S_{y,i}\right)\\
    &=\alpha\beta\left(\frac{S_{+,0}}{2}\sum_{i=1}^{N-1}(S_{x,i}-iS_{y,i})-\frac{S_{-,0}}{2}\sum_{i=1}^{N-1}(S_{x,i}+iS_{y,i})\right)\\
    &=\alpha\beta\left(\frac{S_{+,0}}{2}\sum_{i=1}^{N-1}S_{-,i}-\frac{S_{-,0}}{2}\sum_{i=1}^{N-1}S_{+,i}\right),\\
 \end{aligned}   
\end{equation}
where $\alpha=i\pi(1-e_\mathrm{z}),\beta=-iJ\frac{t}{m},S_{\pm,i}=S_{x,i}\pm i S_{y,i} $

So the exponent of the evolution operator after the application of $m$ H2I pulses can be written as
\begin{equation}
\begin{aligned}
    &-i\tilde{H}t=i\pi mS_{z,0}(1-e_\mathrm{z})\\
    &-iJt\left(\sum_{i=1}^{N-1}S_{x,0}S_{x,i}+S_{y,0}S_{y,i} +S_{z,0}S_{z,i}\right)\\
    &+\pi(1-e_\mathrm{z})J\frac{t}{2}\left(\frac{S_{+,0}}{2}\sum_{i=1}^{N-1}S_{-,i}-\frac{S_{-,0}}{2}\sum_{i=1}^{N-1}S_{+,i}\right)\\
    &=i\pi mS_{z,0}(1-e_\mathrm{z})-iJt\sum_{i=1}^{N-1}S_{z,0}S_{z,i}\\
    &+\left(\pi(1-e_\mathrm{z})J\frac{t}{2}-iJt\right)\frac{S_{+,0}}{2}\sum_{i=1}^{N-1}S_{-,i}\\
     &-\left(\pi(1-e_\mathrm{z})J\frac{t}{2}+iJt\right)\frac{S_{-,0}}{2}\sum_{i=1}^{N-1}S_{+,i}\\
\end{aligned}    
\end{equation}

\begin{figure}
    \includegraphics[width=\columnwidth]{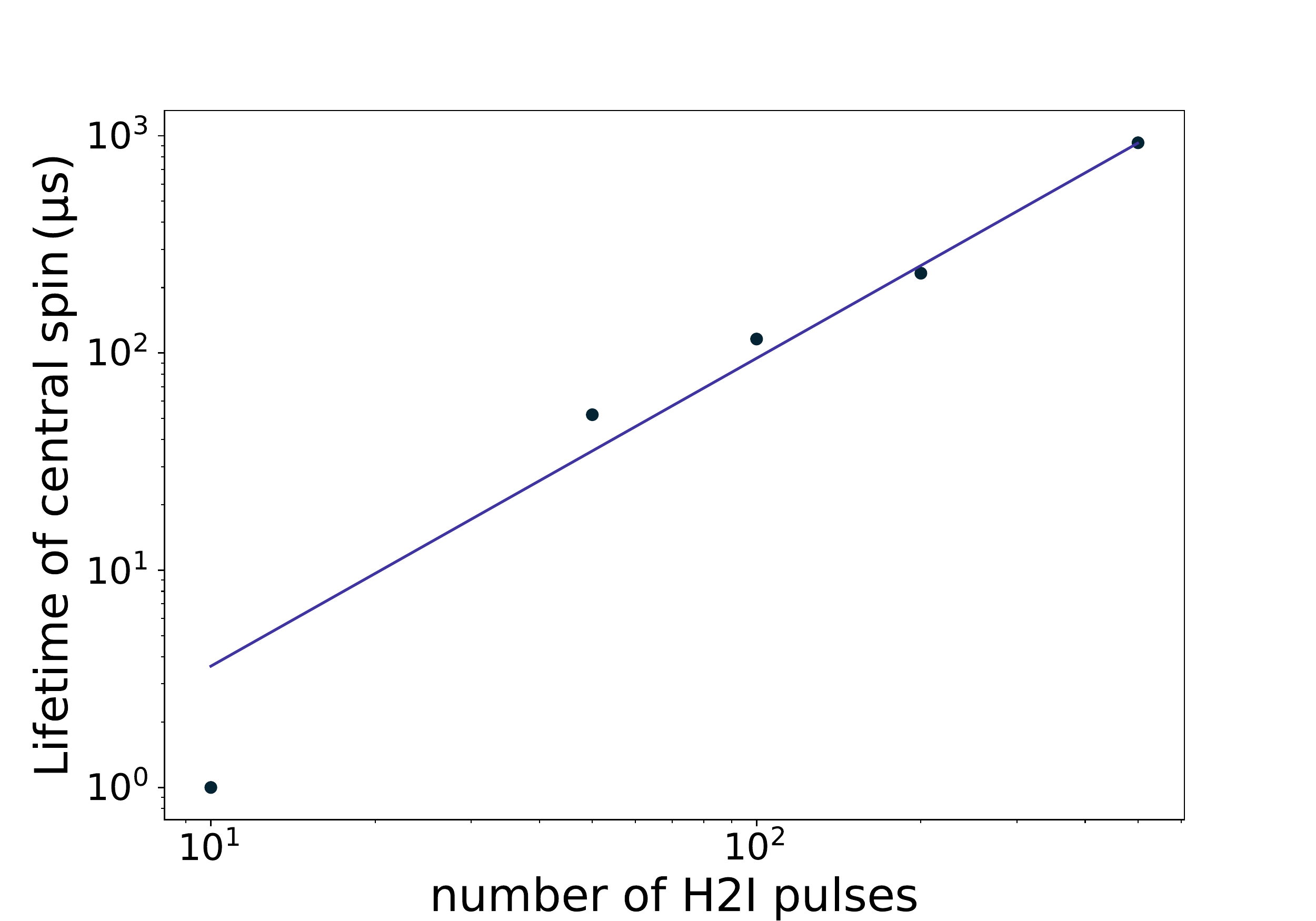}    
    \caption{The number of Floquet cycles (or driving periods) where the stroboscopic central spin magnetization remains $\ge$0.46 versus the number of H2I pulses.  We assume isotropic interactions of strength $J_z/2\pi=J_{xy}/2\pi=1$ MHz. Here, $\delta J/2\pi=0.2$ MHz, $e_\mathrm{c,sat}=0.05$, $e_\mathrm{z}=0.01$, $T=1~\mu$s.}
    \label{fig:h2iscaling}
\end{figure}
We see that the application of $m$ H2I pulses is roughly equivalent to an effective magnetic field on the central spin that is proportional to $m$. The remaining terms in the effective Hamiltonian $\tilde H$ include a longitudinal coupling $J$ and transverse couplings that depend on the error of the H2I pulse. Similarly to having a large Zeeman energy $\rm{B_c}$ on the central spin, the application of sufficiently many H2I pulses stabilizes $z$-basis product states by suppressing in-plane interactions.  

Next, we investigate numerically how the time scale on which the central spin magnetization remains stable depends on $m$. In Fig.~\ref{fig:h2iscaling}, we show that the magnetization increases nonlinearly and polynomially as we increase the number of H2I pulses per driving period. Specifically, the figure shows the number of Floquet cycles over which the stroboscopic central spin magnetization remains $\ge0.46$. We fit the data to the function $(\frac{x}{\alpha})^{\beta}$ and find that $\beta=1.41761895$.

\bibliography{ref}
\end{document}